\documentclass[a4paper,pra,twocolumn,english,notitlepage,nofootinbib,aps,superscriptaddress,showkeys]{revtex4}

\usepackage{amsmath,amsthm,amssymb,graphicx,xcolor,float}
\usepackage{esint}
\usepackage[unicode=true]{hyperref}
\hypersetup{
     colorlinks=true,
     linkcolor=blue,
     citecolor=red,
     urlcolor=magenta,
}
\usepackage{mathrsfs,mathtools,mathdots,dsfont}
\usepackage{listings}
\usepackage{subfigure}

\newtheorem*{theorem*}{Theorem}

\newtheorem*{corollary*}{Corollary}

\newtheorem*{lemma*}{Lemma}

\newtheorem*{proposition*}{Example*}

\newtheorem*{conjecture*}{Conjecture}
\theoremstyle{definition}

\newtheorem*{definition*}{Definition}
\theoremstyle{remark}
\newtheorem{remark}{Remark}
\newtheorem*{remark*}{Remark}

\begin{document}

%\title{Exact SU(2) Yang--Mills Waves from a Simple Ansatz: Three Families of Exact Solutions}

%\title{Exact SU(2) Yang--Mills Waves from a Simple Ansatz: Linear, Nonlinear Self-Interacting, and Pure-Gauge Families}

\title{Exact SU(2) Yang--Mills Waves from a Simple Ansatz}

\author{Yu-Xuan Zhang}
\affiliation{School of Physics, Nankai University, Tianjin 300071, People's Republic of China}

\author{Jing-Ling Chen}
\email{chenjl@nankai.edu.cn}
\affiliation{Theoretical Physics Division, Chern Institute of Mathematics, Nankai University, Tianjin 300071, People's Republic of China}

\date{\today}

\begin{abstract}
We use an ansatz based on rotations of the Pauli basis to construct exact propagating solutions of the sourceless SU(2) Yang--Mills equations in $(3+1)$ dimensions. By examining and extending this ansatz, we obtain two wave families with nontrivial transverse structures and a family of finite-amplitude waves on a self-similar non-Abelian background. These solutions can retain a genuinely non-Abelian structure while admitting arbitrary wave profiles propagating at the speed of light and carrying the corresponding longitudinal electric or magnetic fields. We also investigate the parameter conditions for their non-Abelian character and their energy transport. For the self-similar family, we study the structure of its singular plane and construct suitable sources to replace the singular region while preserving wave propagation in the remaining regions. This offers a possible interpretation of these solutions as propagating fields outside sources.
\end{abstract}


\maketitle

%\tableofcontents

%\newpage

\section{Introduction}

Since its introduction in 1954, Yang--Mills (YM) theory has become a foundation of the standard model of particle physics, describing the strong and electroweak interactions through non-Abelian gauge groups~\cite{YangMills1954}. Unlike Maxwell theory, YM theory contains interactions among the gauge fields themselves, making the classical field equations inherently nonlinear. This self-interaction makes exact, time-dependent propagating solutions difficult to construct, while also making them an important means of understanding non-Abelian field dynamics. Explicit analytic field configurations can also provide reference solutions for numerical studies of real-time evolution.

Early studies of exact solutions in YM theory and its extensions coupled to matter focused mainly on static or spherically symmetric configurations, leading to important results such as magnetic monopoles~\cite{WuYang1969,tHooft1974} and dyons~\cite{JuliaZee1975}. In Euclidean space, the self-duality condition led to instanton solutions~\cite{Belavin1975}, which profoundly influenced our understanding of quantum tunneling and vacuum structure. Extended configurations such as flux tubes and vortex lines were also studied~\cite{NielsenOlesen1973}. These developments revealed the rich spatial structures that gauge-field self-interactions can support and established exact solutions as an important tool for studying nonperturbative physics.

For propagating solutions, Ishida and Hosoya obtained nonlinear periodic waves expressed in terms of Jacobi elliptic functions~\cite{IshidaHosoya1975}. Coleman constructed non-Abelian generalizations of electromagnetic plane waves~\cite{Coleman1977}. Kovacs and Lo studied complex self-dual and anti-self-dual propagating waves, including a lightlike branch with arbitrary phase profiles~\cite{KovacsLo1979}, while Baseyan, Matinyan, and Savvidy obtained nonlinear plane waves with timelike wave vectors~\cite{Baseyan1979}. G\"uven also studied non-Abelian waves coupled to gravity~\cite{Guven1979}. Oh and Teh, together with their collaborators obtained non-Abelian progressive waves and investigated their color radiation~\cite{OhTeh1985,Oh1986PRD}, while Rabinowitch developed constructions of expanding waves, transverse progressive waves, and propagating exterior fields of sources~\cite{Rabinowitch2008,Rabinowitch2025,Rabinowitch2026}. These results include both arbitrary light-speed profiles and nonlinear periodic waves, exhibiting propagation patterns beyond the simple electromagnetic plane-wave picture~\cite{Coleman1977,OhTeh1985}. Meanwhile, the analysis of full plane-wave symmetry by Basler and H\"adicke shows that the non-Abelian character of a propagating solution is restricted by the imposed symmetry conditions~\cite{BaslerHadicke1984,AdamoIlderton2019}. The search for new propagating configurations therefore deepens our understanding of the structure of the gauge vacuum and calls for the identification of suitable transverse structures.

Recently, we constructed and classified a class of complex plane-wave solutions in SU(2) YM theory, obtaining genuinely non-Abelian waves with a nonlinear second harmonic and a zero-action solution family connecting the vacuum to non-Abelian waves~\cite{ZhangChenNPB2026}. Following this line of research, we now seek real-valued wave solutions with nontrivial transverse structure. We investigate the wave profiles that these fields can support and the role of longitudinal fields in non-Abelian energy transport, thereby examining the relation between self-interaction and wave propagation in different spatial configurations.

In recent years, synthetic gauge fields in ultracold atoms~\cite{Lin2009Nature,Lin2009PRL,Goldman2014} and artificial gauge potentials in photonic systems~\cite{Ozawa2019} have provided new experimental means of studying gauge couplings. Engineering effective gauge potentials allows matter-wave or optical propagation to be studied in controlled settings, including schemes for realizing non-Abelian couplings~\cite{Goldman2014}. These developments increase the demand for gauge-field models with explicit spacetime structure. Analytic configurations supplied by exact solutions can serve as starting points for studying matter responses in prescribed gauge backgrounds, complementing studies of the dynamics of the classical YM field itself.

We first construct gauge potentials in a rotated Pauli basis that depends on a transverse coordinate, and extend the basic ansatz by allowing its coefficients to vary with the transverse coordinates and the wave phase. This yields two transverse wave families, for which we give the potentials, field strengths, and conditions for Abelian reduction. In both families, the wave profiles can be arbitrary functions of the light-speed phase $\theta=k(z-ct)$, while the transverse dependence is expressed in terms of parabolic cylinder functions and their degenerate limits. The solutions can also carry a nonzero longitudinal magnetic field. We then derive the energy density and flux, showing that the energy stored in the longitudinal magnetic field does not contribute to the longitudinal flux and explaining the role of transverse flux in the energy balance of phase-dependent backgrounds.

Building on this construction, we relax the time and longitudinal coordinate dependence of the potentials to obtain a self-similar non-Abelian background with both longitudinal electric and magnetic fields. We find six independent propagating modes on this background, each admitting an arbitrary phase profile. These give a family of finite-amplitude waves that allows us to compare the energy transport of the background and the propagating part. We analyze the field invariants, non-Abelian character, and planar singularity of this family, and show that the singularity cannot be removed by choosing the wave profiles. We then join two exterior regions away from the singular plane, determine the surface charge and current supporting the resulting field, and calculate the energy exchange between the source and the field. We also compare the solution families with earlier waves and relevant nonexistence theorems to clarify their propagation properties and range of applicability.

The paper is organized as follows. In Sec.~\ref{sec:II}, we review the vacuum YM equations and introduce the color electric and magnetic fields. In Sec.~\ref{sec:III}, we introduce the transversely dependent rotated Pauli basis, formulate the basic ansatz, and discuss its constant-coefficient solutions. In Sec.~\ref{sec:IV}, we solve the field equations for the extended ansatz, obtain the two transverse wave families and their Abelian conditions, and analyze their energy transport. On this basis, Sec.~\ref{sec:V} further allows a longitudinal electric field, constructs the self-similar background and its propagating waves, and discusses their physical properties. Section~\ref{sec:VI} summarizes the solution families, examines the singularity and surface-source construction, and compares the results with known solutions and relevant nonexistence theorems. Finally, Sec.~\ref{sec:VII} presents our conclusions and outlook.

\section{YM equations in vacuum} \label{sec:II}

We work in Minkowski spacetime with metric $(+,-,-,-)$ and keep $c$ explicitly. The SU(2) gauge potential is $A_\mu = (\varphi, -\vec{A})$, where $\varphi$ is the scalar potential and $\vec{A} = (A_x, A_y, A_z)$ is matrix-valued vector field (in the Lie algebra of SU(2)). The field strength tensor is
\begin{equation}
F_{\mu\nu} = \partial_\mu A_\nu - \partial_\nu A_\mu + {\rm i} g [A_\mu, A_\nu],
\end{equation}
with $g$ the coupling constant. The sourceless YM equations are
\begin{subequations}
\begin{eqnarray}
&&D_\mu F^{\mu\nu} \equiv \partial_\mu F^{\mu\nu} + {\rm i} g [A_\mu, F^{\mu\nu}] = 0, \label{eq:YM1}\\
&& D_\mu F_{\nu\gamma} + D_\nu F_{\gamma\mu} + D_\gamma F_{\mu\nu} = 0, \label{eq:Bianchi}
\end{eqnarray}
\end{subequations}
where the covariant derivative is $D_\mu = \partial_\mu + {\rm i} g A_\mu$. Equation~\eqref{eq:Bianchi} is the Bianchi identity and is automatically satisfied (it reduces to the Jacobi identity). Therefore, to find exact solutions we only need to solve Eq.~\eqref{eq:YM1}.

It is convenient to introduce the color electric and magnetic fields~\cite{Jackson1999,Actor1979}:
\begin{align}
\vec{E} &= -\frac{1}{c}\frac{\partial \vec{A}}{\partial t} - \vec{\nabla}\varphi - {\rm i} g [\varphi, \vec{A}], \label{eq:Edef}\\
\vec{B} &= \vec{\nabla}\times\vec{A} - {\rm i} g (\vec{A}\times\vec{A}), \label{eq:Bdef}
\end{align}
where $(\vec{A}\times\vec{A})_i = \epsilon_{ijk} A_j A_k$, $\epsilon_{ijk}$ is the Levi-Civita symbol,  and the product is matrix multiplication. In terms of these, the YM equation ~\eqref{eq:YM1} becomes
%\begin{subequations}
\begin{eqnarray}
&&\vec{\nabla}\cdot\vec{E} - {\rm i} g\bigl(\vec{A}\cdot\vec{E} - \vec{E}\cdot\vec{A}\bigr) = 0, \label{eq:Gauss}\\
&& -\frac{1}{c}\frac{\partial \vec{E}}{\partial t} + \vec{\nabla}\times\vec{B} - {\rm i} g\bigl([\varphi,\vec{E}] + \vec{A}\times\vec{B} + \vec{B}\times\vec{A}\bigr) = 0, \nonumber\\ \label{eq:Ampere}
\end{eqnarray}
%\end{subequations}
which are the generalized Gauss's law and Amp\`ere's law, respectively. The remaining two equations obtained from Eq.~\eqref{eq:Bianchi} (the analogs of Faraday's law and the absence of magnetic monopoles) follow from the Bianchi identity and need not be imposed separately.

\section{Rotated basis and the ansatz}\label{sec:III}

To find wave solutions, we first make a gauge transformation of the
vacuum by using a color-transport matrix with a wave phase. The resulting
potentials have a wave-like form, although their field strengths vanish.
We then change the coefficients of these potentials to construct an
ansatz for nonzero fields.

Under a local SU(2) transformation $U$, the potentials transform as
\begin{align}
\varphi'&=U\varphi U^{-1}
 +\frac{{\rm i}}{gc}(\partial_tU)U^{-1},\label{eq:III-gaugephi}\\
\vec A'&=U\vec A U^{-1}
 -\frac{{\rm i}}{g}(\vec\nabla U)U^{-1}.\label{eq:III-gaugeA}
\end{align}
Starting from $\varphi=0$ and $\vec A=0$, these expressions give a pure
gauge solution.

Let $\vec\sigma=(\sigma_x,\sigma_y,\sigma_z)$ be the Pauli matrices.
We introduce a $y$-dependent rotated Pauli basis
\begin{align}
\Sigma_x&=\cos(\lambda y)\sigma_x+\sin(\lambda y)\sigma_y,\nonumber\\
\Sigma_y&=-\sin(\lambda y)\sigma_x+\cos(\lambda y)\sigma_y,\nonumber\\
\Sigma_z&=\sigma_z,
\end{align}
where $\lambda$ is an arbitrary real constant (so $\lambda y$ is
dimensionless). These $\Sigma$ matrices obey the same commutation relations:
\begin{align}
[\Sigma_x,\Sigma_y]&=2{\rm i}\Sigma_z,\nonumber\\
[\Sigma_y,\Sigma_z]&=2{\rm i}\Sigma_x,\qquad
[\Sigma_z,\Sigma_x]=2{\rm i}\Sigma_y.\label{eq:sigma}
\end{align}
With $\theta=kz-\omega t$, where $k$ is the wave number and $\omega$ is the
frequency, we choose
\begin{equation}
U={\rm e}^{{\rm i}\theta\Sigma_x/2}
 =\cos\frac\theta2\,I+{\rm i}\Sigma_x\sin\frac\theta2.
\label{eq:III-transport}
\end{equation}
The time and longitudinal derivatives act on $\theta$, whereas the
transverse derivative acts on $\Sigma_x$. Using
$\partial_y\Sigma_x=\lambda\Sigma_y$ and
$\Sigma_y\Sigma_x=-{\rm i}\Sigma_z$, we obtain
\begin{align}
(\partial_tU)U^{-1}&=-\frac{{\rm i}\omega}{2}\Sigma_x,\nonumber\\
(\partial_zU)U^{-1}&=\frac{{\rm i}k}{2}\Sigma_x,\nonumber\\
(\partial_yU)U^{-1}&=\frac{{\rm i}\lambda}{2}
 \bigl[\sin\theta\,\Sigma_y+(\cos\theta-1)\Sigma_z\bigr].
\end{align}
Thus the pure gauge potentials are
\begin{align}
\varphi^{(0)}&=\frac{\omega}{2gc}\Sigma_x,\nonumber\\
\vec A^{(0)}&=\frac{k}{2g}\Sigma_x\vec e_z
 +\frac{\lambda}{2g}
 \bigl[\sin\theta\,\Sigma_y+(\cos\theta-1)\Sigma_z\bigr]\vec e_y.
\label{eq:III-puregauge}
\end{align}
These potentials depend on the wave phase $\theta$, but their field
strengths are $\vec E=\vec B=0$ for arbitrary $k$ and $\omega$.
The constant, sine, and cosine terms in the transverse potential have
thus been obtained from a single gauge transformation.

\begin{remark}\emph{A simple five-parameter ansatz.}
We first replace the fixed coefficients in Eq.~\eqref{eq:III-puregauge}
by five independent real constants. The gauge potentials become
\begin{align}
\varphi&=\alpha_1\Sigma_x,\nonumber\\
\vec A&=\alpha_2\Sigma_x\vec e_z
 +\bigl[(\alpha_3+\alpha_5\cos\theta)\Sigma_z
       +\alpha_4\sin\theta\,\Sigma_y\bigr]\vec e_y.
\label{eq:III-five}
\end{align}
For $g,k,\omega\ne0$, substituting this ansatz into the Gauss law and
the longitudinal component of the Amp\`ere law gives
$k\alpha_1=(\omega/c)\alpha_2$ for a wave with nonzero field strengths.
The transverse component then gives $\omega^2=c^2k^2$.
Taking $\omega=ck$, we have $\alpha_1=\alpha_2$, and the remaining
algebraic equations are
\begin{align}
&\alpha_1\bigl[(\lambda+2g\alpha_3)^2+4g^2\alpha_4^2\bigr]
 -2gk\alpha_4\alpha_5=0,\nonumber\\
&(\lambda+2g\alpha_3)(4g\alpha_1\alpha_5-k\alpha_4)=0,\nonumber\\
&\alpha_1(\alpha_4^2-\alpha_5^2)=0.
\label{eq:III-reduced}
\end{align}
For $\alpha_4\ne0$, these equations give the three branches
\begin{equation}
\begin{array}{c|ccc}
 &\alpha_1=\alpha_2&\alpha_3+\lambda/(2g)&\alpha_5\\ \hline
({\rm a})&0&0&0\\
({\rm b})&\eta k/(4g)&\xi\alpha_4&\eta\alpha_4\\
({\rm c})&\eta k/(2g)&0&\eta\alpha_4
\end{array}
\label{eq:III-branches}
\end{equation}
where $\eta,\xi=\pm1$ and $\alpha_4$ is free. Branch (c) is a pure gauge
solution; without imposing $\omega=ck$, it extends to
$\alpha_1=\eta\omega/(2gc)$ and $\alpha_2=\eta k/(2g)$.
For $\alpha_4=0$, the solutions with nonzero field strengths have
$\alpha_1=\alpha_2=0$ and arbitrary $\alpha_3,\alpha_5$;
the remaining cases have zero field strengths.

We can see that a simple change of the coefficients already gives a
new wave solution. In particular, the fields of branch (b) are
\begin{align}
\vec E&=E_y\vec e_y,\qquad \vec B=-E_y\vec e_x,\nonumber\\
E_y&=\frac{k\alpha_4}{2}
 \bigl[(\cos\theta-\xi\eta)\Sigma_y-\eta\sin\theta\,\Sigma_z\bigr].
\label{eq:III-simplewave}
\end{align}
However, this solution is still a $U(1)$ embedding. A simple criterion
is the existence of a traceless Hermitian color matrix $N$, with
$N^2=I$, satisfying
\begin{align}
\frac1c\frac{\partial N}{\partial t}
 +{\rm i}g[\varphi,N]&=0,\nonumber\\
\vec\nabla N-{\rm i}g[\vec A,N]&=0.
\label{eq:III-U1criterion}
\end{align}
For Eq.~\eqref{eq:III-simplewave}, we take
\begin{equation}
N=\frac{E_y}{\sqrt{\tfrac12{\rm Tr}(E_y^2)}}
\end{equation}
on a region where $E_y\ne0$. Direct substitution verifies
Eq.~\eqref{eq:III-U1criterion}, so the solution is locally a $U(1)$
embedding. The same criterion holds with $N=\Sigma_y$ for branch (a),
and with $N=\Sigma_z$ for the nonzero-field solutions with $\alpha_4=0$. $\blacksquare$
\end{remark}

Based on the above ansatz, we now consider how to obtain more general
solutions. The two Pauli terms in each rotated matrix share one
coefficient. For example,
\begin{align}
\alpha_1\Sigma_x
 &=\alpha_1\cos(\lambda y)\sigma_x
   +\alpha_1\sin(\lambda y)\sigma_y,\nonumber\\
\alpha_4\sin\theta\,\Sigma_y
 &=-\alpha_4\sin\theta\sin(\lambda y)\sigma_x+\alpha_4\sin\theta\cos(\lambda y)\sigma_y.
\end{align}
We keep these rotation factors, but allow their coefficients to vary
independently with $x$ and the wave phase $\theta=kz-\omega t$.
The enlarged ansatz is
\begin{align}
\varphi&=a_1\cos(\lambda y)\sigma_x
        +a_2\sin(\lambda y)\sigma_y,\nonumber\\
\vec A&=\bigl[b_1\cos(\lambda y)\sigma_x
             +b_2\sin(\lambda y)\sigma_y\bigr]\vec e_z\nonumber\\
 &\quad+\bigl[-c_1\sin(\lambda y)\sigma_x
             +c_2\cos(\lambda y)\sigma_y+c_3\sigma_z\bigr]\vec e_y,
\label{eq:III-enlarged}
\end{align}
where $a_1,a_2,b_1,b_2,c_1,c_2,c_3$ are real functions of $(\theta,x)$.
The potentials in Eq.~\eqref{eq:III-five} are recovered by setting
\begin{align}
a_1=a_2&=\alpha_1,\qquad b_1=b_2=\alpha_2,\nonumber\\
c_1=c_2&=\alpha_4\sin\theta,\nonumber\\
c_3&=\alpha_3+\alpha_5\cos\theta.
\end{align}
Thus the five-parameter ansatz is included as a special case. The
additional $x$ dependence also allows a longitudinal magnetic field.
We next calculate the fields to see how this component can couple to
the wave.

\onecolumngrid
\section{Transverse wave solutions}\label{sec:IV}

To find wave solutions of Eq.~\eqref{eq:III-enlarged}, we first compare
the electric and magnetic fields. Their relations lead to a transverse
ansatz whose coefficient equations can be solved explicitly. We take
$g$, $\lambda$ and $k$ to be real and nonzero throughout this section.

\subsection{Reduction of the YM Equations}

Since $\theta=kz-\omega t$ and $A_x=0$, the field definitions give
\begin{align}
E_x&=-\partial_x\varphi,
 & B_x&=\partial_yA_z-k\partial_\theta A_y-{\rm i}g[A_y,A_z],\nonumber\\
E_y&=\frac\omega c\partial_\theta A_y-\partial_y\varphi-{\rm i}g[\varphi,A_y],
 & B_y&=-\partial_xA_z,\nonumber\\
E_z&=\frac\omega c\partial_\theta A_z-k\partial_\theta\varphi-{\rm i}g[\varphi,A_z],
 & B_z&=\partial_xA_y.
\label{eq:IV-field-components}
\end{align}
The commutators containing $\varphi$ are
\begin{align}
[\varphi,A_y]
 &=\varphi A_y-A_y\varphi \nonumber\\
 &=\bigl[a_1c_2\cos^2(\lambda y)
        +a_2c_1\sin^2(\lambda y)\bigr]
        [\sigma_x,\sigma_y] +a_1c_3\cos(\lambda y)[\sigma_x,\sigma_z]
       +a_2c_3\sin(\lambda y)[\sigma_y,\sigma_z] \nonumber\\
 &=2{\rm i}\Bigl\{
       a_2c_3\sin(\lambda y)\sigma_x
       -a_1c_3\cos(\lambda y)\sigma_y +\bigl[a_1c_2\cos^2(\lambda y)
              +a_2c_1\sin^2(\lambda y)\bigr]\sigma_z
       \Bigr\}, \nonumber\\
[\varphi,A_z]
 &=\varphi A_z-A_z\varphi \nonumber\\
 &=(a_1b_2-a_2b_1)\sin(\lambda y)\cos(\lambda y)
       [\sigma_x,\sigma_y] \nonumber\\
 &=2{\rm i}(a_1b_2-a_2b_1)
       \sin(\lambda y)\cos(\lambda y)\sigma_z.
\label{eq:IV-commutators}
\end{align}
The expression for $[A_y,A_z]$ follows by replacing $a_1,a_2$ with
$b_1,b_2$ in $[\varphi,A_y]$ and changing the sign. In particular,
the electric and magnetic components satisfy
\begin{align}
B_x+E_y&=\left(\frac\omega c-k\right)\partial_\theta A_y
 +\partial_y(A_z-\varphi)-{\rm i}g[A_y,A_z-\varphi],\nonumber\\
B_y-E_x&=-\partial_x(A_z-\varphi).
\end{align}
We therefore take
\begin{equation}
\omega=ck,\qquad b_1=a_1,\qquad b_2=a_2,
\qquad \theta=kz-ckt.
\label{eq:IV-alignment}
\end{equation}
This choice leaves five coefficient functions and gives
\begin{equation}
A_z=\varphi,\qquad E_z=0,\qquad B_x=-E_y,\qquad B_y=E_x.
\label{eq:IV-wave-relations}
\end{equation}
These relations define the transverse ansatz considered below.
Substituting Eq.~\eqref{eq:IV-alignment} into the field components gives
\begin{align}
E_x&=-(\partial_xa_1)\cos(\lambda y)\sigma_x
     -(\partial_xa_2)\sin(\lambda y)\sigma_y,\nonumber\\
E_y&=(\lambda a_1-k\partial_\theta c_1+2gc_3a_2)
       \sin(\lambda y)\sigma_x
     -(\lambda a_2-k\partial_\theta c_2+2gc_3a_1)
       \cos(\lambda y)\sigma_y\nonumber\\
 &\quad+\bigl[k\partial_\theta c_3
              +2ga_1c_2\cos^2(\lambda y)+2ga_2c_1\sin^2(\lambda y)\bigr]\sigma_z,
 \nonumber\\
B_z&=-(\partial_xc_1)\sin(\lambda y)\sigma_x
     +(\partial_xc_2)\cos(\lambda y)\sigma_y
     +(\partial_xc_3)\sigma_z.
\label{eq:IV-solution-fields}
\end{align}

With Eq.~\eqref{eq:IV-wave-relations}, the derivative terms in the
Amp\`ere law are
\begin{align}
-\frac1c\frac{\partial\vec E}{\partial t}
 &=k\partial_\theta E_x\,\vec e_x
   +k\partial_\theta E_y\,\vec e_y,\nonumber\\
\vec\nabla\times\vec B
 &=(\partial_yB_z-\partial_zE_x)\vec e_x
   -(\partial_zE_y+\partial_xB_z)\vec e_y
   +(\partial_xE_x+\partial_yE_y)\vec e_z\nonumber\\
 &=(\partial_yB_z-k\partial_\theta E_x)\vec e_x
   -(k\partial_\theta E_y+\partial_xB_z)\vec e_y
   +(\partial_xE_x+\partial_yE_y)\vec e_z.
\end{align}
Thus the phase derivatives cancel in
$-c^{-1}\partial_t\vec E+\vec\nabla\times\vec B$.
The matrix products in the nonlinear terms give
\begin{align}
\vec A\times\vec B+\vec B\times\vec A
 &=(A_yB_z-\varphi E_x+E_x\varphi-B_zA_y)\vec e_x
   +(-\varphi E_y+E_y\varphi)\vec e_y
   +(A_yE_y-E_yA_y)\vec e_z\nonumber\\
 &=\bigl([A_y,B_z]-[\varphi,E_x]\bigr)\vec e_x
   -[\varphi,E_y]\vec e_y+[A_y,E_y]\vec e_z.
\end{align}
Adding $[\varphi,\vec E]=[\varphi,E_x]\vec e_x+
[\varphi,E_y]\vec e_y$ leaves
\begin{equation}
[\varphi,\vec E]+\vec A\times\vec B+\vec B\times\vec A
 =[A_y,B_z]\vec e_x+[A_y,E_y]\vec e_z.
\end{equation}
The three components of the Amp\`ere law are therefore
\begin{equation}
\partial_xB_z=0,\qquad
\partial_yB_z-{\rm i}g[A_y,B_z]=0,\qquad
\partial_xE_x+\partial_yE_y-{\rm i}g[A_y,E_y]=0.
\label{eq:IV-reduced-fields}
\end{equation}
The Gauss law gives the same last equation, since
\begin{equation}
\vec\nabla\cdot\vec E
 -{\rm i}g(\vec A\cdot\vec E-\vec E\cdot\vec A)
 =\partial_xE_x+\partial_yE_y-{\rm i}g(A_yE_y-E_yA_y).
\end{equation}
The remaining two YM equations follow from the Bianchi identity for
the fields constructed from the potentials.

The first equation in Eq.~\eqref{eq:IV-reduced-fields} is
\begin{equation}
0=\partial_xB_z
 =-(\partial_x^2c_1)\sin(\lambda y)\sigma_x
   +(\partial_x^2c_2)\cos(\lambda y)\sigma_y
   +(\partial_x^2c_3)\sigma_z,
\end{equation}
which gives $\partial_x^2c_1=\partial_x^2c_2=\partial_x^2c_3=0$.
For the second equation, the derivative and commutator are
\begin{align}
\partial_yB_z
 &=-\lambda(\partial_xc_1)\cos(\lambda y)\sigma_x
   -\lambda(\partial_xc_2)\sin(\lambda y)\sigma_y,\nonumber\\
[A_y,B_z]
 &=(c_2\partial_xc_3-c_3\partial_xc_2)
       \cos(\lambda y)[\sigma_y,\sigma_z]\nonumber\\
 &\quad+(c_1\partial_xc_3-c_3\partial_xc_1)
       \sin(\lambda y)[\sigma_z,\sigma_x]\nonumber\\
 &\quad+(c_2\partial_xc_1-c_1\partial_xc_2)
       \sin(\lambda y)\cos(\lambda y)[\sigma_x,\sigma_y].
\end{align}
Using the Pauli commutators, their combination becomes
\begin{align}
0={}&\bigl[-\lambda\partial_xc_1
       +2g(c_2\partial_xc_3-c_3\partial_xc_2)\bigr]
       \cos(\lambda y)\sigma_x\nonumber\\
 &+\bigl[-\lambda\partial_xc_2
       +2g(c_1\partial_xc_3-c_3\partial_xc_1)\bigr]
       \sin(\lambda y)\sigma_y\nonumber\\
 &+2g(c_2\partial_xc_1-c_1\partial_xc_2)
       \sin(\lambda y)\cos(\lambda y)\sigma_z.
\label{eq:IV-magnetic-expansion}
\end{align}
Each Pauli component must vanish.

For the Gauss equation, differentiation of
Eq.~\eqref{eq:IV-solution-fields} gives
\begin{align}
\partial_xE_x+\partial_yE_y
 &=
 \bigl[-\partial_x^2a_1
       +\lambda(\lambda a_1-k\partial_\theta c_1+2gc_3a_2)\bigr]
       \cos(\lambda y)\sigma_x\nonumber\\
 &\quad+
 \bigl[-\partial_x^2a_2
       +\lambda(\lambda a_2-k\partial_\theta c_2+2gc_3a_1)\bigr]
       \sin(\lambda y)\sigma_y\nonumber\\
 &\quad-4g\lambda(a_1c_2-a_2c_1)
       \sin(\lambda y)\cos(\lambda y)\sigma_z.
\label{eq:IV-Gauss-divergence}
\end{align}
In particular, the last term comes from
\begin{equation}
\partial_y\bigl[
2ga_1c_2\cos^2(\lambda y)+2ga_2c_1\sin^2(\lambda y)\bigr]
 =-4g\lambda(a_1c_2-a_2c_1)\sin(\lambda y)\cos(\lambda y).
\end{equation}
The commutator can be expanded as
\begin{equation}
[A_y,E_y]
 =-c_1\sin(\lambda y)[\sigma_x,E_y]
  +c_2\cos(\lambda y)[\sigma_y,E_y]
  +c_3[\sigma_z,E_y].
\end{equation}
Substituting the three Pauli components of $E_y$ gives
\begin{align}
-{\rm i}g[A_y,E_y]
 &=2g\Bigl\{
 c_3(\lambda a_2-k\partial_\theta c_2+2gc_3a_1)
 +c_2\bigl[k\partial_\theta c_3
       +2ga_1c_2\cos^2(\lambda y)+2ga_2c_1\sin^2(\lambda y)\bigr]
 \Bigr\}\cos(\lambda y)\sigma_x\nonumber\\
 &\quad+2g\Bigl\{
 c_3(\lambda a_1-k\partial_\theta c_1+2gc_3a_2)
 +c_1\bigl[k\partial_\theta c_3
       +2ga_1c_2\cos^2(\lambda y)+2ga_2c_1\sin^2(\lambda y)\bigr]
 \Bigr\}\sin(\lambda y)\sigma_y\nonumber\\
 &\quad+2g\Bigl\{
 c_1(\lambda a_2-k\partial_\theta c_2+2gc_3a_1)
 -c_2(\lambda a_1-k\partial_\theta c_1+2gc_3a_2)
 \Bigr\}\sin(\lambda y)\cos(\lambda y)\sigma_z.
\label{eq:IV-Gauss-commutator}
\end{align}
Adding the $\sigma_z$ terms in
Eqs.~\eqref{eq:IV-Gauss-divergence} and
\eqref{eq:IV-Gauss-commutator}, and taking the coefficient of
$\sin(\lambda y)\cos(\lambda y)$, yields
\begin{align}
0={}&-4g\lambda(a_1c_2-a_2c_1)\nonumber\\
 &+2g\bigl[
 c_1(\lambda a_2-k\partial_\theta c_2+2gc_3a_1)
 -c_2(\lambda a_1-k\partial_\theta c_1+2gc_3a_2)
 \bigr]\nonumber\\
={}&2g\bigl[
 -3\lambda(a_1c_2-a_2c_1)
 +2gc_3(c_1a_1-c_2a_2)
 +k(c_2\partial_\theta c_1-c_1\partial_\theta c_2)
 \bigr].
\label{eq:IV-Gauss-expansion}
\end{align}

The $\sigma_x$ and $\sigma_y$ components also contain third harmonics.
Their quadratic trigonometric factor is
\begin{equation}
a_1c_2\cos^2(\lambda y)+a_2c_1\sin^2(\lambda y)
 =\frac{a_1c_2+a_2c_1}{2}
  +\frac{a_1c_2-a_2c_1}{2}\cos(2\lambda y).
\end{equation}
Using
\begin{align}
\cos(\lambda y)\cos(2\lambda y)
 &=\tfrac12[\cos(3\lambda y)+\cos(\lambda y)],\nonumber\\
\sin(\lambda y)\cos(2\lambda y)
 &=\tfrac12[\sin(3\lambda y)-\sin(\lambda y)],
\end{align}
the third-harmonic part of the Gauss equation is
\begin{equation}
g^2c_2(a_1c_2-a_2c_1)\cos(3\lambda y)\sigma_x
 +g^2c_1(a_1c_2-a_2c_1)\sin(3\lambda y)\sigma_y=0.
\end{equation}
Thus
\begin{equation}
c_2(a_1c_2-a_2c_1)=0,\qquad
c_1(a_1c_2-a_2c_1)=0.
\end{equation}
Multiplying these equations by $a_1$ and $-a_2$, respectively, and
adding gives $(a_1c_2-a_2c_1)^2=0$. Hence
\begin{equation}
a_1c_2=a_2c_1.
\label{eq:IV-harmonic-constraint}
\end{equation}
The quadratic factor consequently reduces to $a_1c_2=a_2c_1$,
independent of $y$. The coefficients of
$\cos(\lambda y)\sigma_x$ and $\sin(\lambda y)\sigma_y$ in the
Gauss equation are then
\begin{align}
\partial_x^2a_1
 &=\lambda(\lambda a_1-k\partial_\theta c_1+2gc_3a_2)
   +2gc_3(\lambda a_2-k\partial_\theta c_2+2gc_3a_1)
   +2gc_2(k\partial_\theta c_3+2ga_1c_2),\nonumber\\
\partial_x^2a_2
 &=\lambda(\lambda a_2-k\partial_\theta c_2+2gc_3a_1)
   +2gc_3(\lambda a_1-k\partial_\theta c_1+2gc_3a_2)
   +2gc_1(k\partial_\theta c_3+2ga_2c_1).
\end{align}
Equation~\eqref{eq:IV-harmonic-constraint} also removes the first
term in the square brackets of Eq.~\eqref{eq:IV-Gauss-expansion}.
Collecting these results, the coefficient equations are
\begin{gather}
\partial_x^2c_1=\partial_x^2c_2=\partial_x^2c_3=0,\nonumber\\
\lambda\partial_xc_1=2g(c_2\partial_xc_3-c_3\partial_xc_2),\qquad
\lambda\partial_xc_2=2g(c_1\partial_xc_3-c_3\partial_xc_1),\nonumber\\
c_2\partial_xc_1-c_1\partial_xc_2=0,
\label{eq:IV-c-equations}\\
a_1c_2=a_2c_1,\qquad
2gc_3(c_1a_1-c_2a_2)
 +k(c_2\partial_\theta c_1-c_1\partial_\theta c_2)=0,
\label{eq:IV-algebraic}
\end{gather}
\begin{align}
\partial_x^2a_1
 &=[\lambda^2+4g^2(c_3^2+c_2^2)]a_1+4g\lambda c_3a_2
+k\bigl(
 2gc_2\partial_\theta c_3-\lambda\partial_\theta c_1
 -2gc_3\partial_\theta c_2\bigr),\nonumber\\
\partial_x^2a_2
 &=[\lambda^2+4g^2(c_3^2+c_1^2)]a_2+4g\lambda c_3a_1
+k\bigl(
 2gc_1\partial_\theta c_3-\lambda\partial_\theta c_2
 -2gc_3\partial_\theta c_1\bigr).
\label{eq:IV-a-equations}
\end{align}

\twocolumngrid

\subsection{Solutions of the Equations}\label{sec:IV-solutions}

We now solve Eqs.~\eqref{eq:IV-c-equations}--\eqref{eq:IV-a-equations}
for two families of wave solutions. We integrate in $x$ with $\theta$
held fixed, so the integration constants are functions of $\theta$.
Functions with the same names in the two families are chosen independently.

\subsubsection{Solution family I}

We first take $c_1=c_2=0$, so that $A_y=c_3\sigma_z$.
The algebraic equations \eqref{eq:IV-algebraic} are then satisfied,
and \mbox{Eq.}~\eqref{eq:IV-c-equations} gives
\begin{equation}
 \partial_x^2c_3=0,\qquad
 c_3=c_0(\theta)+c_4(\theta)x,
 \label{eq:IV-longitudinal-background}
\end{equation}
where $c_0$ and $c_4$ are arbitrary real functions.
The two equations for $a_1,a_2$ become
\begin{align}
 \partial_x^2a_1
 &=(\lambda^2+4g^2c_3^2)a_1+4g\lambda c_3a_2,\nonumber\\
 \partial_x^2a_2
 &=(\lambda^2+4g^2c_3^2)a_2+4g\lambda c_3a_1.
 \label{eq:IV-coupled}
\end{align}
Their sum and difference give independent equations. We therefore set
\begin{gather}
 h_\pm=a_1\pm a_2,\nonumber\\
 \partial_x^2h_\pm=(\lambda\pm2gc_3)^2h_\pm.
 \label{eq:IV-decoupled}
\end{gather}
The original amplitudes are recovered from
$a_1=(h_++h_-)/2$ and $a_2=(h_+-h_-)/2$.
We specify the two integration functions for each mode by its value
and $x$ derivative at $x=0$:
\begin{align}
 F_\pm(\theta)&=h_\pm(\theta,0),\nonumber\\
 G_\pm(\theta)&=\partial_xh_\pm(\theta,0).
 \label{eq:IV-independent-data}
\end{align}

For $c_4\ne0$, introduce
\begin{equation}
 \zeta_\pm=\sqrt{4|gc_4|}
 \left(x+\frac{2gc_0\pm\lambda}{2gc_4}\right).
 \label{eq:IV-zeta-independent}
\end{equation}
Then $\partial_x^2=4|gc_4|\partial_{\zeta_\pm}^2$ and
$(\lambda\pm2gc_3)^2=|gc_4|\zeta_\pm^2$, so
\mbox{Eq.}~\eqref{eq:IV-decoupled} becomes
\begin{equation}
 \partial_{\zeta_\pm}^2h_\pm-\frac{\zeta_\pm^2}{4}h_\pm=0.
 \label{eq:IV-cylinder-equation}
\end{equation}
The parabolic cylinder function $D_\nu(\zeta)$ satisfies
$\partial_\zeta^2D_\nu+
(\nu+\tfrac12-\zeta^2/4)D_\nu=0$.
Thus the order is $\nu=-1/2$, and two independent solutions are
\begin{align}
 U_\pm(x)&=D_{-1/2}(\zeta_\pm(x)),\nonumber\\
 V_\pm(x)&=D_{-1/2}(-\zeta_\pm(x)).
 \label{eq:IV-independent-basis}
\end{align}
Their dependence on $\theta$ is understood through $c_0,c_4$.
Primes on $U_\pm,V_\pm$ denote $x$ derivatives. Using their Wronskian
$U_\pm V_\pm'-U_\pm'V_\pm=2\sqrt{2|gc_4|}$, the solution with
the values in \mbox{Eq.}~\eqref{eq:IV-independent-data} is
\begin{align}
 h_\pm(\theta,x)
 &=\frac{F_\pm V_\pm'(0)-G_\pm V_\pm(0)}
         {2\sqrt{2|gc_4|}}\,U_\pm(x)\nonumber\\
 &\quad+\frac{G_\pm U_\pm(0)-F_\pm U_\pm'(0)}
                {2\sqrt{2|gc_4|}}\,V_\pm(x).
 \label{eq:IV-independent-modes}
\end{align}

For $c_4=0$, \mbox{Eq.}~\eqref{eq:IV-decoupled} has constant coefficients,
and the same integration functions give
\begin{align}
 h_\pm
 &=F_\pm\cosh\bigl[(\lambda\pm2gc_0)x\bigr]\nonumber\\
 &\quad+G_\pm\frac{\sinh\bigl[(\lambda\pm2gc_0)x\bigr]}
              {\lambda\pm2gc_0}.
 \label{eq:IV-independent-constant}
\end{align}
When $\lambda\pm2gc_0=0$, the corresponding expression is
$h_\pm=F_\pm+G_\pm x$.

The six functions $c_0,c_4,F_\pm,G_\pm$ specify this family.
Substituting the amplitudes into the ansatz gives the potentials
\begin{align}
 \varphi=A_z
 &=\frac{h_++h_-}{2}\cos(\lambda y)\sigma_x\nonumber\\
 &\quad+\frac{h_+-h_-}{2}\sin(\lambda y)\sigma_y,\nonumber\\
 A_x&=0,\qquad A_y=(c_0+c_4x)\sigma_z.
 \label{eq:IV-independent-potentials}
\end{align}
The field components follow from \mbox{Eq.}~\eqref{eq:IV-solution-fields}:
\begin{align}
 E_x
 &=-\frac12\bigl[
   (\partial_xh_++\partial_xh_-)\cos(\lambda y)\sigma_x\nonumber\\
 &\qquad+(\partial_xh_+-\partial_xh_-)\sin(\lambda y)\sigma_y
  \bigr],\nonumber\\
 E_y
 &=\frac12\bigl[
   (\lambda+2gc_3)h_+\nonumber\\
 &\qquad+(\lambda-2gc_3)h_-
  \bigr]\sin(\lambda y)\sigma_x\nonumber\\
 &\quad-\frac12\bigl[
   (\lambda+2gc_3)h_+\nonumber\\
 &\qquad-(\lambda-2gc_3)h_-
  \bigr]\cos(\lambda y)\sigma_y\nonumber\\
 &\quad+k(\partial_\theta c_0+x\partial_\theta c_4)\sigma_z,\nonumber\\
 E_z&=0,\qquad B_z=c_4\sigma_z,\nonumber\\
 B_x&=-E_y,\qquad B_y=E_x.
 \label{eq:IV-independent-fields}
\end{align}
Here $c_3=c_0+c_4x$, and the $x$ derivatives of $h_\pm$ are obtained
directly from Eqs.~\eqref{eq:IV-independent-modes} and
\eqref{eq:IV-independent-constant}.

\subsubsection{Solution family II}

We next take $c_2=\varepsilon c_1$, where $\varepsilon=\pm1$.
The first equation in \mbox{Eq.}~\eqref{eq:IV-algebraic} then reads
$c_1(a_2-\varepsilon a_1)=0$. We take $a_2=\varepsilon a_1$,
which also satisfies the second equation.
The equations for $c_1,c_3$ reduce to
\begin{gather}
 \partial_x^2c_1=\partial_x^2c_3=0,\nonumber\\
 \left(c_3+\frac{\varepsilon\lambda}{2g}\right)\partial_xc_1
 -c_1\partial_xc_3=0.
 \label{eq:IV-proportional-background}
\end{gather}
The last equation fixes the direction of the pair
$(c_1,c_3+\varepsilon\lambda/(2g))$ as $x$ varies.
We therefore introduce a real amplitude $R(\theta,x)$ and an angle
$\chi(\theta)$, and write
\begin{gather}
 c_1=R\sin\chi,\qquad c_2=\varepsilon R\sin\chi,\nonumber\\
 c_3=R\cos\chi-\frac{\varepsilon\lambda}{2g}.
 \label{eq:IV-proportional-parameters}
\end{gather}
The second-derivative equations give $\partial_x^2R=0$, hence
\begin{equation}
 R=c_0(\theta)+c_4(\theta)x,
 \label{eq:IV-proportional-amplitude}
\end{equation}
with arbitrary real functions $c_0,c_4$.

After substituting these coefficients into \mbox{Eq.}~\eqref{eq:IV-a-equations},
the two equations for $a_1,a_2$ coincide.
The coefficient of $a_1$ becomes $4g^2R^2$, while the phase-derivative
terms simplify using
\begin{equation}
 c_1\partial_\theta c_3
 -\left(c_3+\frac{\varepsilon\lambda}{2g}\right)\partial_\theta c_1
 =-R^2\partial_\theta\chi.
\end{equation}
The resulting equation is
\begin{equation}
 \partial_x^2a_1
 =4g^2R^2a_1-2g\varepsilon kR^2\partial_\theta\chi.
 \label{eq:IV-angle-equation}
\end{equation}
Since $\chi$ depends only on $\theta$,
$\varepsilon k\partial_\theta\chi/(2g)$ is an $x$-independent
particular solution. We subtract it by writing
\begin{equation}
 a_1=\frac{\varepsilon k}{2g}\partial_\theta\chi+h,\qquad
 a_2=\varepsilon a_1.
 \label{eq:IV-proportional-electric}
\end{equation}
The remaining amplitude $h(\theta,x)$ satisfies
\begin{equation}
 \partial_x^2h=4g^2(c_0+c_4x)^2h.
 \label{eq:IV-angle-homogeneous}
\end{equation}
As in family I, we specify its two integration functions at $x=0$:
\begin{equation}
 F(\theta)=h(\theta,0),\qquad
 G(\theta)=\partial_xh(\theta,0).
 \label{eq:IV-proportional-data}
\end{equation}

For $c_4\ne0$, set
\begin{gather}
 \zeta=\sqrt{4|gc_4|}\left(x+\frac{c_0}{c_4}\right),\nonumber\\
 U(x)=D_{-1/2}(\zeta(x)),\quad
 V(x)=D_{-1/2}(-\zeta(x)).
 \label{eq:IV-proportional-basis}
\end{gather}
Equation~\eqref{eq:IV-angle-homogeneous} then becomes
$\partial_\zeta^2h-\zeta^2h/4=0$, the same parabolic cylinder
equation as in family I. With primes again denoting $x$ derivatives,
its solution is
\begin{align}
 h(\theta,x)
 &=\frac{FV'(0)-GV(0)}{2\sqrt{2|gc_4|}}\,U(x)\nonumber\\
 &\quad+\frac{GU(0)-FU'(0)}{2\sqrt{2|gc_4|}}\,V(x).
 \label{eq:IV-proportional-mode}
\end{align}
For $c_4=0$, the solution is
\begin{equation}
 h=F\cosh(2gc_0x)+G\frac{\sinh(2gc_0x)}{2gc_0},
 \label{eq:IV-proportional-constant}
\end{equation}
with $h=F+Gx$ when $c_0=0$.

The five functions $c_0,c_4,\chi,F,G$ specify this family.
The potentials are
\begin{align}
 A_x&=0,\qquad A_z=\varphi,\nonumber\\
 \varphi
 &=\left(\frac{\varepsilon k}{2g}\partial_\theta\chi+h\right)
   \bigl[\cos(\lambda y)\sigma_x
        +\varepsilon\sin(\lambda y)\sigma_y\bigr],\nonumber\\
 A_y
 &=R\sin\chi\,
   \bigl[-\sin(\lambda y)\sigma_x
         +\varepsilon\cos(\lambda y)\sigma_y\bigr]\nonumber\\
 &\quad+\left(R\cos\chi-\frac{\varepsilon\lambda}{2g}\right)\sigma_z.
 \label{eq:IV-proportional-potentials}
\end{align}
On substitution into \mbox{Eq.}~\eqref{eq:IV-solution-fields}, the terms
proportional to $\partial_\theta\chi$ cancel in $E_y$.
The resulting fields are
\begin{align}
 E_x
 &=-(\partial_xh)
   \bigl[\cos(\lambda y)\sigma_x
        +\varepsilon\sin(\lambda y)\sigma_y\bigr],\nonumber\\
 E_y
 &=\bigl[2g\varepsilon Rh\cos\chi
         -k(\partial_\theta R)\sin\chi\bigr]\nonumber\\
 &\quad\times\bigl[\sin(\lambda y)\sigma_x
        -\varepsilon\cos(\lambda y)\sigma_y\bigr]\nonumber\\
 &\quad+\bigl[k(\partial_\theta R)\cos\chi
              +2g\varepsilon Rh\sin\chi\bigr]\sigma_z,\nonumber\\
 E_z&=0,\qquad B_x=-E_y,\qquad B_y=E_x,\nonumber\\
 B_z
 &=c_4\bigl[-\sin\chi\sin(\lambda y)\sigma_x\nonumber\\
 &\qquad+\varepsilon\sin\chi\cos(\lambda y)\sigma_y
           +\cos\chi\,\sigma_z\bigr],
 \label{eq:IV-proportional-fields}
\end{align}
where $\partial_\theta R=\partial_\theta c_0+x\partial_\theta c_4$.
The derivative $\partial_xh$ follows from
Eqs.~\eqref{eq:IV-proportional-mode} and
\eqref{eq:IV-proportional-constant}.

\begin{remark}\emph{Overlap of the families and the $U(1)$ wave.}
For $\chi\equiv0$, family II is a single mode of family I.
The identification is obtained by replacing $c_0$ in
\mbox{Eq.}~\eqref{eq:IV-independent-potentials} with
$c_0-\varepsilon\lambda/(2g)$ and taking
\begin{equation}
 h_+=(1+\varepsilon)h,\qquad h_-=(1-\varepsilon)h.
\end{equation}
The $U(1)$ wave in \mbox{Eq.}~\eqref{eq:III-simplewave} is also
contained in family II. For a real constant $\alpha_4$, choose
\begin{gather}
 \varepsilon=1,\qquad c_4=F=G=0,\nonumber\\
 \chi=\frac{\theta}{2},\qquad
 c_0=2\alpha_4\cos\frac{\theta}{2}.
 \label{eq:IV-five-parameter-recovery}
\end{gather}
This is branch (b) of
\mbox{Eq.}~\eqref{eq:III-branches}, with \mbox{$\xi=\eta=1$}.
\hfill\(\blacksquare\)
\end{remark}

\begin{remark}\emph{Single-color and pure-gauge solutions.}
For $c_1^2\ne c_2^2$, the coefficient equations give
$\partial_xc_1=\partial_xc_2=\partial_xc_3=0$.
The nonzero-field solutions have one fixed color direction.
With arbitrary real functions $F(\theta),G(\theta)$, let
\begin{equation}
 h=F\cosh(\lambda x)+G\frac{\sinh(\lambda x)}{\lambda}.
 \label{eq:IV-single-color-mode}
\end{equation}
The two choices are
\begin{align}
 c_2=c_3=a_2&=0,\qquad
 a_1=\frac{k}{\lambda}\partial_\theta c_1+h,
 \nonumber\\
 c_1=c_3=a_1&=0,\qquad
 a_2=\frac{k}{\lambda}\partial_\theta c_2+h.
 \label{eq:IV-single-color-solutions}
\end{align}
The surviving $c_1$ or $c_2$ is an arbitrary real function of
$\theta$, and the functions in the two choices are independent.
Both are $U(1)$ embeddings. Their fields coincide with family I
at $c_0=c_4=0$, with $h_+=h_-=h$ or $h_+=-h_-=h$,
respectively; the additional terms containing $c_1$ or $c_2$
contribute only pure-gauge freedom.

For the pure-gauge solutions with $c_1^2\ne c_2^2$,
let $\rho(\theta)>0$ be arbitrary,
$\varepsilon=\pm1$, and let $w(\rho)$ be a real nonvanishing
Painlev\'e III function satisfying
\begin{equation}
 \frac{\mathrm d^2w}{\mathrm d\rho^2}
 =\frac1w\left(\frac{\mathrm dw}{\mathrm d\rho}\right)^2
  -\frac1\rho\frac{\mathrm dw}{\mathrm d\rho}
  -\frac{w^3}{4}+\frac1{4w}.
 \label{eq:IV-flat-Painleve}
\end{equation}
With $w$ evaluated at $\rho(\theta)$, the coefficients are
\begin{align}
 c_1&=\frac{\lambda\rho}{4g}
          \left(w+\frac{\varepsilon}{w}\right),\qquad
 c_2=\frac{\lambda\rho}{4g}
          \left(w-\frac{\varepsilon}{w}\right),\nonumber\\
 c_3&=\frac{\lambda\rho}{2gw}\frac{\mathrm dw}{\mathrm d\rho},
 \nonumber\\
 a_1&=\frac{k\partial_\theta\rho}{\lambda\rho}c_1,\qquad
 a_2=\frac{k\partial_\theta\rho}{\lambda\rho}c_2.
 \label{eq:IV-flat-coefficients}
\end{align}
These connections are locally gauge-equivalent to the vacuum.
The flat cases with $c_1^2=c_2^2$ are included in
families I and II: both require constant $c_0$ and $c_4=0$,
with $G_\pm=0$, $(\lambda\pm2gc_0)F_\pm=0$ in family I,
and $G=0$, $c_0F=0$ in family II.
\hfill\(\blacksquare\)
\end{remark}

\subsection{Abelian Character and Energy Transport}
\label{sec:IV-properties}

We first examine whether the solutions are $U(1)$ embeddings.
According to Eq.~\eqref{eq:III-U1criterion}, this requires a traceless
Hermitian color matrix $N$, with $N^2=I$, satisfying
\begin{gather}
 \partial_xN=0,\qquad
 \partial_yN-{\rm i}g[A_y,N]=0,\nonumber\\
 k\partial_\theta N-{\rm i}g[\varphi,N]=0.
 \label{eq:IV-property-color}
\end{gather}
Here we have used $A_x=0$ and $A_z=\varphi$. These conditions also
imply that $N$ commutes with every field component. We then calculate
the energy density and flux with the normalization
${\rm Tr}(\sigma_i\sigma_j)=2\delta_{ij}$:
\begin{align}
 \mathcal E&=\frac12{\rm Tr}(\vec E\cdot\vec E+\vec B\cdot\vec B),
 \nonumber\\
 \vec{\mathcal S}&=c\,{\rm Tr}(\vec E\times\vec B).
 \label{eq:IV-property-energy}
\end{align}

\subsubsection{Solution family I}

For family I, $A_y=c_3\sigma_z$ and $B_z=c_4\sigma_z$, where
$c_3=c_0+c_4x$. When $c_4\ne0$, the condition $[N,B_z]=0$ fixes
$N=\pm\sigma_z$. The last equation in
Eq.~\eqref{eq:IV-property-color} then gives
$[\varphi,\sigma_z]=0$, hence $a_1=a_2=0$.
An Abelian embedding with $c_4\ne0$ therefore requires
$h_+=h_-=0$, or equivalently
$F_\pm=G_\pm=0$. This choice also works for arbitrary $c_0,c_4$:
all nonzero fields lie along $\sigma_z$. In the presence of
$c_4\ne0$, either nonzero electric mode produces a non-Abelian field.

When $c_4=0$, further Abelian reductions occur. For $E_x\ne0$,
its color direction must agree with $N$. Substituting this direction
into the $y$ equation in Eq.~\eqref{eq:IV-property-color} gives
\begin{align}
 0={}&\lambda(\partial_xa_1)(\partial_xa_2)\nonumber\\
 &+2gc_0(\partial_xa_1)^2\cos^2(\lambda y)\nonumber\\
 &+2gc_0(\partial_xa_2)^2\sin^2(\lambda y).
 \label{eq:IV-I-color-condition}
\end{align}
Comparing the constant and $\cos(2\lambda y)$ terms and using
the phase equation selects $a_2=0$, $N=\sigma_x$, or
$a_1=0$, $N=\sigma_y$, when $c_0=0$. For $c_0\ne0$, it gives
\begin{gather}
 c_0=-\frac{\varepsilon\lambda}{2g},\qquad
 a_2=\varepsilon a_1,\qquad \varepsilon=\pm1,\nonumber\\
 N=\cos(\lambda y)\sigma_x+\varepsilon\sin(\lambda y)\sigma_y.
 \label{eq:IV-I-resonant-color}
\end{gather}
The amplitude equation now reduces to $\partial_x^2a_1=0$.
Although this color direction varies with $y$, it is covariantly
constant and therefore also defines a $U(1)$ embedding. If $E_x$
vanishes identically, the amplitude equations leave $\varphi=0$
or a zero-field member of Eq.~\eqref{eq:IV-I-resonant-color}.
These cases, together with $h_+=h_-=0$, give the Abelian reductions
of family I.

For the energy calculation, we use
$a_1=(h_++h_-)/2$, $a_2=(h_+-h_-)/2$ and the fields in
Eq.~\eqref{eq:IV-independent-fields}. Since $E_z=0$,
$B_x=-E_y$ and $B_y=E_x$, substitution gives
\begin{align}
 \mathcal E
 &=c_4^2+{\rm Tr}(E_x^2+E_y^2)\nonumber\\
 &=c_4^2+2k^2(\partial_\theta c_3)^2\nonumber\\
 &\quad+2(\partial_xa_1)^2\cos^2(\lambda y)
       +2(\partial_xa_2)^2\sin^2(\lambda y)\nonumber\\
 &\quad+2(\lambda a_1+2gc_3a_2)^2\sin^2(\lambda y)\nonumber\\
 &\quad+2(\lambda a_2+2gc_3a_1)^2\cos^2(\lambda y).
 \label{eq:IV-I-energy}
\end{align}

Only the $\sigma_z$ component of $E_y$ contributes to the transverse
flux: ${\rm Tr}(E_yB_z)=2kc_4\partial_\theta c_3$, whereas
${\rm Tr}(E_xB_z)=0$. Thus
\begin{align}
 \vec{\mathcal S}
 &=c\,{\rm Tr}(E_yB_z)\vec e_x
   -c\,{\rm Tr}(E_xB_z)\vec e_y\nonumber\\
 &\quad+c\,{\rm Tr}(E_x^2+E_y^2)\vec e_z\nonumber\\
 &=2ckc_4(\partial_\theta c_3)\vec e_x
   +c(\mathcal E-c_4^2)\vec e_z.
 \label{eq:IV-I-flux}
\end{align}
Here $\partial_\theta c_3=\partial_\theta c_0+x\partial_\theta c_4$.
The electric modes $h_\pm$ determine the additional longitudinal
energy flux; the transverse flux is controlled entirely by
$c_0,c_4$. In particular, $c_4=0$ gives
$\vec{\mathcal S}=c\mathcal E\vec e_z$ for both Abelian and
non-Abelian members of the family.

\subsubsection{Solution family II}

The fields in Eq.~\eqref{eq:IV-proportional-fields} give
\begin{equation}
 -\frac12{\rm Tr}[E_x,E_y]^2
 =4(\partial_xh)^2\bigl[4g^2R^2h^2+k^2(\partial_\theta R)^2\bigr].
 \label{eq:IV-II-commutator}
\end{equation}
For a nonzero background $R=c_0+c_4x$, the equation
$\partial_x^2h=4g^2R^2h$ prevents a nonzero solution $h$ from
being constant in $x$. The field commutator is then nonzero in
a region, so the solution is non-Abelian. Thus $h$ supplies
the non-Abelian part of this family.

When $h=0$, the fields are $E_x=0$,
$E_y=k(\partial_\theta R)N$ and $B_z=c_4N$, with their common
color direction given by
\begin{align}
 N={}&-\sin\chi\sin(\lambda y)\sigma_x\nonumber\\
    &+\varepsilon\sin\chi\cos(\lambda y)\sigma_y
      +\cos\chi\,\sigma_z.
 \label{eq:IV-II-color}
\end{align}
This is the matrix $N$ required by Eq.~\eqref{eq:III-U1criterion}.
Indeed, $N^2=I$ and $A_y=RN-\varepsilon\lambda\sigma_z/(2g)$
give $\partial_xN=0$ and
$\partial_yN-{\rm i}g[A_y,N]=0$. The remaining derivative is
\begin{align}
 &k\partial_\theta N-{\rm i}g[\varphi,N]\nonumber\\
 &\quad=2g\varepsilon h\cos\chi\sin(\lambda y)\sigma_x\nonumber\\
 &\qquad-2gh\cos\chi\cos(\lambda y)\sigma_y\nonumber\\
 &\qquad+2g\varepsilon h\sin\chi\,\sigma_z,
 \label{eq:IV-II-color-residual}
\end{align}
which vanishes for $h=0$. Hence $F=G=0$ gives a $U(1)$
embedding for arbitrary $c_0,c_4,\chi$.

The degenerate background $c_0=c_4=0$ also allows nonzero $h$.
Here $h=F+Gx$, $E_y=B_z=0$, and the color direction in
Eq.~\eqref{eq:IV-I-resonant-color} satisfies the Abelian criterion.
This is the resonant member shared with family I.
\par

For the energy and flux, the Pauli traces of
Eq.~\eqref{eq:IV-proportional-fields} give
\begin{align}
 \frac12{\rm Tr}E_x^2
 &=(\partial_xh)^2,\qquad
 \frac12{\rm Tr}B_z^2=c_4^2,\nonumber\\
 \frac12{\rm Tr}E_y^2
 &=[2g\varepsilon Rh\cos\chi
       -k(\partial_\theta R)\sin\chi]^2\nonumber\\
 &\quad+[k(\partial_\theta R)\cos\chi
       +2g\varepsilon Rh\sin\chi]^2\nonumber\\
 &=4g^2R^2h^2+k^2(\partial_\theta R)^2,\nonumber\\
 {\rm Tr}(E_xB_z)&=0,\qquad
 {\rm Tr}(E_yB_z)=2kc_4\partial_\theta R.
 \label{eq:IV-II-electric-trace}
\end{align}

The terms proportional to $Rh\,\partial_\theta R$ cancel between
the two squares. Substitution into
Eq.~\eqref{eq:IV-property-energy} yields
\begin{align}
 \mathcal E
 &=c_4^2+2(\partial_xh)^2+8g^2R^2h^2
       +2k^2(\partial_\theta R)^2,\nonumber\\
 \vec{\mathcal S}
 &=2ckc_4(\partial_\theta R)\vec e_x
       +c(\mathcal E-c_4^2)\vec e_z.
 \label{eq:IV-II-energy-flux}
\end{align}
The angle $\chi$ cancels from both quantities. For $h=0$, the
energy is $c_4^2+2k^2(\partial_\theta R)^2$; a nonzero $h$ adds
$2(\partial_xh)^2+8g^2R^2h^2$ and changes only the longitudinal
flux. As in family I, the transverse flux depends on
$\partial_\theta R=\partial_\theta c_0+x\partial_\theta c_4$.

Both families therefore have the same energy-transport relation.
For constant $c_0,c_4$, the transverse flux vanishes, and
$\mathcal S_z=c(\mathcal E-c_4^2)$ separates the constant
longitudinal magnetic energy from the wave energy. The latter
travels along $+z$ at speed $c$, with its profile depending on
$\theta=k(z-ct)$. If $c_4=0$, all the energy contributes to this
longitudinal flux. When $c_4$ varies with the phase, the transverse
flux supplies the balance required by energy conservation:
\begin{align}
 \partial_t\mathcal E+\partial_z\mathcal S_z
 &=-2ckc_4\partial_\theta c_4\nonumber\\
 &=-\partial_x\mathcal S_x.
 \label{eq:IV-energy-conservation}
\end{align}
Since $\mathcal S_y=0$, this is
$\partial_t\mathcal E+\vec\nabla\cdot\vec{\mathcal S}=0$.

\section{Self-similar solutions}\label{sec:V}

The transverse solutions of the preceding section were obtained with
$A_x=0$, $A_z=\varphi$, and the specified Pauli components of
Eq.~\eqref{eq:III-enlarged}. Their dependence on $t,z$ was entirely
through $\theta=kz-ckt$. We now allow $A_x$ and $A_y$ to be arbitrary
color matrices of $(\theta,x,y)$, and allow $\varphi$ to depend
separately on $t$ and $z$, while retaining $A_z=\varphi$.
The latter extension is needed for a longitudinal electric field:
with $A_z=\varphi$, its definition gives
\begin{equation}
 E_z=-\left(\frac1c\partial_t+\partial_z\right)\varphi.
 \label{eq:V-longitudinal-definition}
\end{equation}
A potential depending only on $\theta$ is annihilated by this
derivative. We shall determine the additional dependence allowed
by the YM equations, then use the resulting equations to construct
a self-similar solution.

\subsection{Reduction of the Field Equations}

The enlarged ansatz is
\begin{align}
 \varphi&=\varphi(t,x,y,z),\qquad A_z=\varphi,\nonumber\\
 A_x&=A_x(\theta,x,y),\qquad A_y=A_y(\theta,x,y),
 \label{eq:V-ansatz}
\end{align}
with traceless Hermitian matrix potentials. For $A_x$ and $A_y$,
$c^{-1}\partial_t=-k\partial_\theta$ and
$\partial_z=k\partial_\theta$; these replacements do not yet apply
to $\varphi$. Substitution into the field definitions gives
\begin{align}
 E_x&=k\partial_\theta A_x-\partial_x\varphi
                         -{\rm i}g[\varphi,A_x],\nonumber\\
 E_y&=k\partial_\theta A_y-\partial_y\varphi
                         -{\rm i}g[\varphi,A_y],\nonumber\\
 B_x&=\partial_y\varphi-k\partial_\theta A_y
                         -{\rm i}g[A_y,\varphi]=-E_y,\nonumber\\
 B_y&=k\partial_\theta A_x-\partial_x\varphi
                         -{\rm i}g[\varphi,A_x]=E_x,\nonumber\\
 B_z&=\partial_xA_y-\partial_yA_x-{\rm i}g[A_x,A_y].
 \label{eq:V-fields}
\end{align}
Thus the enlargement preserves $B_x=-E_y$ and $B_y=E_x$.

We first substitute these fields into the Amp\`ere law. The derivative
terms become
\begin{align}
 &-\frac1c\frac{\partial\vec E}{\partial t}
       +\vec\nabla\times\vec B\nonumber\\
 &\quad=\left[-\left(\frac1c\partial_t+\partial_z\right)E_x
                       +\partial_yB_z\right]\vec e_x\nonumber\\
 &\qquad+\left[-\left(\frac1c\partial_t+\partial_z\right)E_y
                       -\partial_xB_z\right]\vec e_y\nonumber\\
 &\qquad+\left(-\frac1c\partial_tE_z
                       +\partial_xE_x+\partial_yE_y\right)\vec e_z.
 \label{eq:V-Ampere-derivatives}
\end{align}
The matrix products in the nonlinear term are
\begin{align}
 &\vec A\times\vec B+\vec B\times\vec A\nonumber\\
 &\quad=(A_yB_z-\varphi E_x+E_x\varphi-B_zA_y)\vec e_x
 \nonumber\\
 &\qquad+(-\varphi E_y-A_xB_z+B_zA_x+E_y\varphi)\vec e_y
 \nonumber\\
 &\qquad+(A_xE_x+A_yE_y-E_yA_y-E_xA_x)\vec e_z.
 \label{eq:V-matrix-products}
\end{align}
Adding $[\varphi,\vec E]$ cancels the terms
$-[\varphi,E_x]$ and $-[\varphi,E_y]$ in the first two
components. The three remaining components are
$[A_y,B_z]$, $-[A_x,B_z]$, and
$[A_x,E_x]+[A_y,E_y]+[\varphi,E_z]$.

These commutators occur together with transverse derivatives.
For a color matrix $M$, we therefore introduce
\begin{align}
 D_xM&=\partial_xM-{\rm i}g[A_x,M],\nonumber\\
 D_yM&=\partial_yM-{\rm i}g[A_y,M].
 \label{eq:V-transverse-derivatives}
\end{align}
Combining the derivative and commutator terms, the Amp\`ere law is
\begin{gather}
 \left(\frac1c\partial_t+\partial_z\right)E_x=D_yB_z,\nonumber\\
 \left(\frac1c\partial_t+\partial_z\right)E_y=-D_xB_z,\nonumber\\
 D_xE_x+D_yE_y-\frac1c\partial_tE_z
                  -{\rm i}g[\varphi,E_z]=0.
 \label{eq:V-Ampere-components}
\end{gather}
The Gauss law gives
\begin{align}
 0&=\partial_xE_x+\partial_yE_y+\partial_zE_z\nonumber\\
 &\quad-{\rm i}g\bigl([A_x,E_x]+[A_y,E_y]+[\varphi,E_z]\bigr)
 \nonumber\\
 &=D_xE_x+D_yE_y+\partial_zE_z-{\rm i}g[\varphi,E_z].
 \label{eq:V-Gauss}
\end{align}
Subtracting the last Amp\`ere equation from the Gauss equation yields
\begin{equation}
 \left(\frac1c\partial_t+\partial_z\right)E_z=0.
 \label{eq:V-Ez-phase}
\end{equation}
Consequently $E_z=E_z(\theta,x,y)$. Equation~
\eqref{eq:V-longitudinal-definition} can now be integrated, because
\begin{align*}
 \left(\frac1c\partial_t+\partial_z\right)\theta=0, \quad
 \left(\frac1c\partial_t+\partial_z\right)\frac{ct+z}{2}=1.
\end{align*}
Denoting the integration matrix by $\Phi(\theta,x,y)$, we obtain
\begin{equation}
 \varphi=\Phi(\theta,x,y)-\frac{ct+z}{2}E_z(\theta,x,y).
 \label{eq:V-affine-potential}
\end{equation}
Thus the YM equations determine the additional longitudinal
dependence: it is linear in $ct+z$, with coefficient fixed by $E_z$.

Substitution into the first two electric components gives
\begin{align}
 E_x&=k\partial_\theta A_x-D_x\Phi+\frac{ct+z}{2}D_xE_z,
 \nonumber\\
 E_y&=k\partial_\theta A_y-D_y\Phi+\frac{ct+z}{2}D_yE_z.
 \label{eq:V-transverse-electric}
\end{align}
All coefficients here depend on $t,z$ only through $\theta$.
Applying $c^{-1}\partial_t+\partial_z$ to these expressions
therefore gives $D_xE_z$ and $D_yE_z$, respectively.
The transverse Amp\`ere equations reduce to
\begin{equation}
 D_xE_z=D_yB_z,\qquad D_yE_z=-D_xB_z.
 \label{eq:V-transverse-equations}
\end{equation}

For the remaining Gauss equation, the separate contributions are
\begin{equation}
 \begin{gathered}
 \begin{aligned}
 D_xE_x+D_yE_y
 &=k(D_x\partial_\theta A_x+D_y\partial_\theta A_y)\\
 &\quad-(D_x^2+D_y^2)\Phi
       +\frac{ct+z}{2}(D_x^2+D_y^2)E_z,
 \end{aligned}\\
 \partial_zE_z=k\partial_\theta E_z,\qquad
 -{\rm i}g[\varphi,E_z]=-{\rm i}g[\Phi,E_z].
 \end{gathered}
 \label{eq:V-Gauss-terms}
\end{equation}
The coefficient of $ct+z$ is already fixed by the transverse equations:
\begin{align}
 (D_x^2+D_y^2)E_z
 &=D_xD_yB_z-D_yD_xB_z\nonumber\\
 &=-{\rm i}g[B_z,B_z]=0.
 \label{eq:V-compatibility}
\end{align}
The remaining terms give the equation for $\Phi$.
The full reduction is therefore
\begin{equation}
 \begin{gathered}
 B_z=\partial_xA_y-\partial_yA_x-{\rm i}g[A_x,A_y],\\
 D_xE_z=D_yB_z,\qquad D_yE_z=-D_xB_z.
 \end{gathered}
 \label{eq:V-reduced-transverse}
\end{equation}
\begin{equation}
 \begin{aligned}
 &(D_x^2+D_y^2)\Phi+{\rm i}g[\Phi,E_z]\\
 &\quad=k(D_x\partial_\theta A_x+D_y\partial_\theta A_y
                          +\partial_\theta E_z).
 \end{aligned}
 \label{eq:V-reduced-profile}
\end{equation}
Together with Eq.~\eqref{eq:V-affine-potential}, these equations
are equivalent to the Gauss and Amp\`ere laws for
Eq.~\eqref{eq:V-ansatz}. The remaining two YM equations follow
from the Bianchi identity.

The structure of the reduced system gives the order of solution.
Equation~\eqref{eq:V-reduced-transverse} determines the transverse
potentials and the longitudinal fields without reference to $\Phi$.
Once they have been found, Eq.~\eqref{eq:V-reduced-profile} is a
linear equation for $\Phi$.
Setting $E_z=0$ and $A_x=0$ returns the field equations
\eqref{eq:IV-reduced-fields}; restricting the Pauli components
as in Eq.~\eqref{eq:IV-alignment} then recovers the preceding
five-function ansatz.

\subsection{Solutions of the Equations}

We first solve Eq.~\eqref{eq:V-reduced-transverse}. For a
self-similar transverse configuration, potentials proportional to
$x^{-1}$ have a useful property: their derivatives and commutators
both scale as $x^{-2}$. We take two different Pauli directions,
\begin{equation}
 A_x=-\frac{p}{2gx}\sigma_z,\qquad
 A_y=-\frac{q}{2gx}\sigma_x,
 \label{eq:V-transverse-trial}
\end{equation}
where $p,q$ are nonzero real constants. We work on $x>0$ and seek
$E_z$ and $\Phi$ independent of $y$. The longitudinal electric field
is still to be determined by the transverse equations.

The derivative and commutator in $B_z$ are
\begin{align*}
 \partial_xA_y&=\frac{q}{2gx^2}\sigma_x,\\
 [A_x,A_y]
 &=\frac{pq}{4g^2x^2}[\sigma_z,\sigma_x]
   =\frac{{\rm i}pq}{2g^2x^2}\sigma_y.
\end{align*}
Consequently,
\begin{equation}
 B_z=\frac{q}{2gx^2}(\sigma_x+p\sigma_y).
 \label{eq:V-magnetic-trial}
\end{equation}
Its covariant derivatives follow from
\begin{align*}
 \partial_xB_z&=-\frac{2q}{2gx^3}(\sigma_x+p\sigma_y),\\
 -{\rm i}g[A_x,B_z]
 &=\frac{pq}{2gx^3}(p\sigma_x-\sigma_y),
\end{align*}
and are
\begin{align}
 D_xB_z&=\frac{q}{2gx^3}
                [(p^2-2)\sigma_x-3p\sigma_y],\nonumber\\
 D_yB_z&=-{\rm i}g[A_y,B_z]
          =-\frac{pq^2}{2gx^3}\sigma_z.
 \label{eq:V-magnetic-derivatives}
\end{align}

Since $E_z$ is independent of $y$, the equation
$D_yE_z=-D_xB_z$ becomes
\begin{equation}
 \frac{{\rm i}q}{2x}[\sigma_x,E_z]
 =-\frac{q}{2gx^3}[(p^2-2)\sigma_x-3p\sigma_y].
 \label{eq:V-electric-direction}
\end{equation}
The commutator on the left has no $\sigma_x$ component, so
$p^2=2$. Comparing the other two components fixes the
$\sigma_z$ coefficient of $E_z$ and sets its $\sigma_y$
coefficient to zero. Thus
\begin{equation}
 E_z=e(\theta,x)\sigma_x+\frac{3p}{2gx^2}\sigma_z,
 \label{eq:V-electric-trial}
\end{equation}
where $e$ is the remaining undetermined coefficient.
The other transverse equation now reads
\begin{align}
 D_xE_z
 &=(\partial_xe)\sigma_x-\frac{pe}{x}\sigma_y
                            -\frac{3p}{gx^3}\sigma_z\nonumber\\
 &=D_yB_z=-\frac{pq^2}{2gx^3}\sigma_z.
 \label{eq:V-transverse-balance}
\end{align}
It gives $\partial_xe=0$, $pe=0$, and $q^2=6$.
Since $p\ne0$, we have $e=0$. Constant color rotations relate
the choices of signs for $p$ and $q$, so we choose
$p=\sqrt2$ and $q=\sqrt6$. The transverse potentials and
longitudinal fields are therefore
\begin{align}
 A_x&=-\frac{\sqrt2}{2gx}\sigma_z,\qquad
 A_y=-\frac{\sqrt6}{2gx}\sigma_x,\nonumber\\
 E_z&=\frac{3\sqrt2}{2gx^2}\sigma_z,\quad
 B_z=\frac{\sqrt6}{2gx^2}(\sigma_x+\sqrt2\sigma_y).
 \label{eq:V-transverse-solution}
\end{align}

All four matrices in Eq.~\eqref{eq:V-transverse-solution} are
independent of $\theta$. The right-hand side of
Eq.~\eqref{eq:V-reduced-profile} therefore vanishes, leaving
\begin{equation}
 (D_x^2+D_y^2)\Phi+{\rm i}g[\Phi,E_z]=0.
 \label{eq:V-profile-equation}
\end{equation}
To solve it, write
\begin{align}
 \Phi&=\Phi_1(\theta,x)\sigma_x
       +\Phi_2(\theta,x)\sigma_y+\Phi_3(\theta,x)\sigma_z.
 \label{eq:V-profile-components}
\end{align}
The first derivatives are
\begin{align}
 D_x\Phi&=\partial_x\Phi
       +\frac{\sqrt2}{x}(\Phi_2\sigma_x-\Phi_1\sigma_y),\nonumber\\
 D_y\Phi&=\frac{\sqrt6}{x}(\Phi_3\sigma_y-\Phi_2\sigma_z).
 \label{eq:V-profile-first-derivatives}
\end{align}
Applying the same derivatives once more gives
\begin{align}
 D_x^2\Phi
 &=\partial_x^2\Phi+\frac{2\sqrt2}{x}
         [(\partial_x\Phi_2)\sigma_x-(\partial_x\Phi_1)\sigma_y]
 \nonumber\\
 &\quad-\frac{\sqrt2}{x^2}(\Phi_2\sigma_x-\Phi_1\sigma_y)
-\frac2{x^2}(\Phi_1\sigma_x+\Phi_2\sigma_y),\nonumber\\
 D_y^2\Phi
 &=-\frac6{x^2}(\Phi_2\sigma_y+\Phi_3\sigma_z),\nonumber\\
 {\rm i}g[\Phi,E_z]
 &=\frac{3\sqrt2}{x^2}(-\Phi_2\sigma_x+\Phi_1\sigma_y).
 \label{eq:V-profile-second-derivatives}
\end{align}
Substitution into Eq.~\eqref{eq:V-profile-equation} and comparison
of the three Pauli components yield
\begin{align}
 x^2\partial_x^2\Phi_1
 &=-2\sqrt2x\partial_x\Phi_2+2\Phi_1+4\sqrt2\Phi_2,\nonumber\\
 x^2\partial_x^2\Phi_2
 &=2\sqrt2x\partial_x\Phi_1-4\sqrt2\Phi_1+8\Phi_2,\nonumber\\
 x^2\partial_x^2\Phi_3&=6\Phi_3.
 \label{eq:V-profile-ODE}
\end{align}
There are no phase derivatives in these equations, so $\theta$
is a parameter in the integration with respect to $x$.

The third equation is independent of the first two.
Substituting $\Phi_3\propto x^m$ gives
$m(m-1)=6$, with roots $m=-2,3$. Hence
\begin{equation}
 \Phi_3=\frac{f_1(\theta)}{x^2}+x^3f_2(\theta),
 \label{eq:V-independent-profile}
\end{equation}
where $f_1,f_2$ are arbitrary real functions.
For the coupled pair, take
$\Phi_1=C_1x^m$ and $\Phi_2=C_2x^m$, with $C_1,C_2$ constant
in $x$. The first two equations in
Eq.~\eqref{eq:V-profile-ODE} become
\begin{align}
 (m^2-m-2)C_1+2\sqrt2(m-2)C_2&=0,\nonumber\\
 -2\sqrt2(m-2)C_1+(m^2-m-8)C_2&=0.
 \label{eq:V-power-equations}
\end{align}
A nonzero pair requires
\begin{align}
 0&=(m^2-m-2)(m^2-m-8)+8(m-2)^2\nonumber\\
  &=(m-2)(m-3)(m^2+3m+8).
 \label{eq:V-power-roots}
\end{align}
The root $m=2$ gives $C_2=0$, whereas $m=3$ gives
$C_2=-\sqrt2C_1$. These yield the two real solutions
\begin{align}
 (\Phi_1,\Phi_2)&=x^2f_3(\theta)(1,0),\quad  x^3f_4(\theta)(1,-\sqrt2).
 \label{eq:V-real-power-modes}
\end{align}
For the remaining roots, Eq.~\eqref{eq:V-power-equations} gives
\begin{equation}
 m=-\frac32\pm\frac{{\rm i}\sqrt{23}}2,\qquad
 \frac{C_2}{C_1}=\frac{1\mp{\rm i}\sqrt{23}}{4\sqrt2}.
 \label{eq:V-complex-power-modes}
\end{equation}
Combining the conjugate modes gives two real solutions oscillating
in $\ln x$. Introduce a fixed length $\ell>0$ to make the
logarithm dimensionless. Together with
Eq.~\eqref{eq:V-independent-profile}, the full real solution is
\begin{align}
 \Phi_1
 &=x^2f_3+x^3f_4+x^{-3/2}f_5
       \cos\left(\frac{\sqrt{23}}2\ln\frac{x}{\ell}\right)\nonumber\\
 &\quad+x^{-3/2}f_6
       \sin\left(\frac{\sqrt{23}}2\ln\frac{x}{\ell}\right),\nonumber\\
 \Phi_2
 &=-\sqrt2x^3f_4+\frac{x^{-3/2}}{4\sqrt2}(f_5-\sqrt{23}f_6)
       \cos\left(\frac{\sqrt{23}}2\ln\frac{x}{\ell}\right)\nonumber\\
 &\quad+\frac{x^{-3/2}}{4\sqrt2}(\sqrt{23}f_5+f_6)
       \sin\left(\frac{\sqrt{23}}2\ln\frac{x}{\ell}\right),\nonumber\\
 \Phi_3&=x^{-2}f_1+x^3f_2.
 \label{eq:V-complete-profile}
\end{align}
Here $f_1,\ldots,f_6$ are arbitrary smooth real functions of
$\theta$. The six independent modes solve the three second-order
equations in Eq.~\eqref{eq:V-profile-ODE} for the transverse
configuration \eqref{eq:V-transverse-solution}.

The potentials now follow from
Eq.~\eqref{eq:V-affine-potential}:
\begin{align}
 \varphi=A_z
 &=\Phi_1\sigma_x+\Phi_2\sigma_y+\left[\Phi_3-\frac{3\sqrt2(ct+z)}{4gx^2}\right]\sigma_z,
 \nonumber\\
 A_x&=-\frac{\sqrt2}{2gx}\sigma_z,\qquad
 A_y=-\frac{\sqrt6}{2gx}\sigma_x.
 \label{eq:V-complete-potentials}
\end{align}
Using Eq.~\eqref{eq:V-transverse-electric}, the fields are
\begin{align}
 E_x
 &=-\left(\partial_x\Phi_1+\frac{\sqrt2}{x}\Phi_2\right)\sigma_x
-\left(\partial_x\Phi_2-\frac{\sqrt2}{x}\Phi_1\right)\sigma_y
 \nonumber\\
 &\quad-\left[\partial_x\Phi_3
                  +\frac{3\sqrt2(ct+z)}{2gx^3}\right]\sigma_z,
 \nonumber\displaybreak[1]\\
 E_y
 &=\left[-\frac{\sqrt6}{x}\Phi_3
                  +\frac{3\sqrt3(ct+z)}{2gx^3}\right]\sigma_y
+\frac{\sqrt6}{x}\Phi_2\sigma_z,\nonumber\displaybreak[1]\\
 E_z&=\frac{3\sqrt2}{2gx^2}\sigma_z,
 \quad B_x=-E_y,\nonumber\\
  B_y&=E_x, \quad
 B_z=\frac{\sqrt6}{2gx^2}(\sigma_x+\sqrt2\sigma_y).
 \label{eq:V-complete-fields}
\end{align}
Setting $f_1=\cdots=f_6=0$ gives a self-similar background:
under $(ct,x,y,z)\mapsto s(ct,x,y,z)$, for $s>0$, its potentials
scale as $s^{-1}$ and its fields as $s^{-2}$.
General phase-dependent profiles need not preserve this symmetry.

\subsection{Abelian Character and Energy Transport}

We first examine the color structure. From
Eq.~\eqref{eq:V-complete-fields}, the longitudinal fields give
\begin{equation}
 \begin{gathered}
 [E_z,B_z]=\frac{3{\rm i}\sqrt3}{g^2x^4}
                         (-\sqrt2\sigma_x+\sigma_y),\\
 -\frac12{\rm Tr}[E_z,B_z]^2=\frac{81}{g^4x^8}.
 \end{gathered}
 \label{eq:V-nonabelian}
\end{equation}
A $U(1)$ embedding requires the field components to commute.
The nonzero commutator therefore excludes such an embedding for
every choice of the six profiles. In particular, the self-similar
background itself is non-Abelian.

The two longitudinal fields have equal color norms and are
orthogonal under the trace:
\begin{gather}
 \frac12{\rm Tr}E_z^2=\frac12{\rm Tr}B_z^2=\frac9{2g^2x^4},
 \nonumber\\
 {\rm Tr}(E_zB_z)=0.
 \label{eq:V-longitudinal-traces}
\end{gather}
Since $B_x=-E_y$ and $B_y=E_x$, it follows that
\begin{equation}
 \begin{gathered}
 {\rm Tr}(\vec E\cdot\vec E-\vec B\cdot\vec B)
       ={\rm Tr}(E_z^2-B_z^2)=0,\\
 {\rm Tr}(\vec E\cdot\vec B)={\rm Tr}(E_zB_z)=0.
 \end{gathered}
 \label{eq:V-quadratic-invariants}
\end{equation}
Thus these quadratic invariants vanish even though the fields
cannot be reduced to a single color direction.

For the energy density, Eq.~\eqref{eq:IV-property-energy} and
the transverse field relations give
\begin{align}
 \mathcal E
 &=\frac12{\rm Tr}(E_z^2+B_z^2)
                     +{\rm Tr}(E_x^2+E_y^2)\nonumber\\
 &=\frac9{g^2x^4}
   +2\biggl(\partial_x\Phi_1+\frac{\sqrt2}{x}\Phi_2\biggr)^2
+2\biggl(\partial_x\Phi_2-\frac{\sqrt2}{x}\Phi_1\biggr)^2
 \nonumber\\
 &\quad+2\biggl(\partial_x\Phi_3
                  +\frac{3\sqrt2(ct+z)}{2gx^3}\biggr)^2
 \nonumber\\
 &\quad+2\biggl(-\frac{\sqrt6}{x}\Phi_3
                  +\frac{3\sqrt3(ct+z)}{2gx^3}\biggr)^2
       +\frac{12}{x^2}\Phi_2^2.
 \label{eq:V-energy}
\end{align}
Here the cross terms between different Pauli components vanish
because $ {\rm Tr}(\sigma_i\sigma_j)=2\delta_{ij}$.

The three flux components are
\begin{align}
 \mathcal S_x&=c\,{\rm Tr}(E_yB_z-E_zE_x),\nonumber\\
 \mathcal S_y&=-c\,{\rm Tr}(E_xB_z+E_zE_y),\nonumber\\
 \mathcal S_z&=c\,{\rm Tr}(E_x^2+E_y^2).
 \label{eq:V-flux-traces}
\end{align}
For the first component, the two nonzero traces are
\begin{align*}
 {\rm Tr}(E_yB_z)
 &=-\frac{6\sqrt2}{gx^3}\Phi_3
                    +\frac{9(ct+z)}{g^2x^5},\\
 {\rm Tr}(E_zE_x)
 &=-\frac{3\sqrt2}{gx^2}\partial_x\Phi_3
                    -\frac{9(ct+z)}{g^2x^5}.
\end{align*}
Evaluating the remaining traces in the same way yields
\begin{align}
 \mathcal S_x
 &=\frac{3\sqrt2c}{gx^2}
       \left(\partial_x\Phi_3-\frac2x\Phi_3\right)
       +\frac{18c(ct+z)}{g^2x^5},\nonumber\\
 \mathcal S_y
 &=\frac{\sqrt6c}{gx^2}
       \left(\partial_x\Phi_1-\frac2x\Phi_1\right)+\frac{2\sqrt3c}{gx^2}
       \left(\partial_x\Phi_2-\frac2x\Phi_2\right),\nonumber\\
 \mathcal S_z&=c\left(\mathcal E-\frac9{g^2x^4}\right).
 \label{eq:V-flux}
\end{align}
The phase derivatives in $\partial_t\mathcal E$ cancel those in
$\partial_z\mathcal S_z$, and $\partial_y\mathcal S_y=0$.
The remaining derivatives give
\begin{equation}
 \begin{aligned}
 &\partial_t\mathcal E+\vec\nabla\cdot\vec{\mathcal S}=\frac{3\sqrt2c}{gx^2}
       \left(\partial_x^2\Phi_3-\frac6{x^2}\Phi_3\right)=0
 \end{aligned}
 \label{eq:V-energy-conservation}
\end{equation}
by Eq.~\eqref{eq:V-profile-ODE}.

We can now separate the self-similar background from a propagating
profile. Setting $\Phi_1=\Phi_2=\Phi_3=0$ in
Eqs.~\eqref{eq:V-energy} and \eqref{eq:V-flux} gives
\begin{align}
 \mathcal E&=\frac9{g^2x^4}
                 +\frac{45(ct+z)^2}{2g^2x^6},\nonumber\\
 \vec{\mathcal S}
 &=\frac{18c(ct+z)}{g^2x^5}\vec e_x
   +\frac{45c(ct+z)^2}{2g^2x^6}\vec e_z.
 \label{eq:V-background-energy}
\end{align}
This background has a transverse energy flux. In fact,
\begin{equation}
 \mathcal E^2-\frac{|\vec{\mathcal S}|^2}{c^2}
 =\frac{81}{g^4x^8}\left[1+\frac{(ct+z)^2}{x^2}\right]>0.
 \label{eq:V-background-flux}
\end{equation}
The vanishing invariants in
Eq.~\eqref{eq:V-quadratic-invariants} therefore do not imply
$|\vec{\mathcal S}|=c\mathcal E$.

For an explicit wave, retain only $f_3(\theta)$ in
Eq.~\eqref{eq:V-complete-profile}, so that
$\Phi=x^2f_3(\theta)\sigma_x$.
The transverse electric fields become
\begin{align}
 E_x
 &=xf_3(-2\sigma_x+\sqrt2\sigma_y)
                 -\frac{3\sqrt2(ct+z)}{2gx^3}\sigma_z,\nonumber\\
 E_y&=\frac{3\sqrt3(ct+z)}{2gx^3}\sigma_y.
 \label{eq:V-wave-fields}
\end{align}
The other fields retain their values in
Eq.~\eqref{eq:V-complete-fields}.
The new part of $E_x$ is trace-orthogonal both to its background
value and to $B_z$, since
\begin{align*}
 {\rm Tr}[(-2\sigma_x+\sqrt2\sigma_y)\sigma_z]&=0,\\
 {\rm Tr}[(-2\sigma_x+\sqrt2\sigma_y)
                    (\sigma_x+\sqrt2\sigma_y)]&=0.
\end{align*}
Consequently the additional energy is $2x^2f_3^2(4+2)=12x^2f_3^2$,
and the full energy and flux are
\begin{align}
 \mathcal E
 &=\frac9{g^2x^4}+\frac{45(ct+z)^2}{2g^2x^6}
                      +12x^2f_3(\theta)^2,\nonumber\\
 \vec{\mathcal S}
 &=\frac{18c(ct+z)}{g^2x^5}\vec e_x+c\left[\frac{45(ct+z)^2}{2g^2x^6}
                      +12x^2f_3(\theta)^2\right]\vec e_z.
 \label{eq:V-wave-energy}
\end{align}
The energy contributed by $f_3(\theta)$ travels along $+z$ at
speed $c$. This is an exact finite-amplitude wave on the
non-Abelian background. A periodic $f_3$ gives a periodic wave,
while compact support in $\theta$ gives a pulse along the
propagation direction.

Finally, Eq.~\eqref{eq:V-nonabelian} identifies $x=0$ as a genuine
curvature singularity. Equation~\eqref{eq:V-energy} also shows
$\mathcal E\ge9/(g^2x^4)$, independently of the wave profiles,
so their choice cannot remove this singularity.
The solution is a vacuum field on a domain away from that plane.
Its energy integral diverges at the plane, and the example in
Eq.~\eqref{eq:V-wave-energy} grows at transverse infinity.
Thus a pulse localized in $\theta$ is not, by itself, a
finite-energy isolated configuration; an exterior interpretation
also requires a source or boundary construction.

\section{Physical Discussion}\label{sec:VI}

We summarize the potentials and their Abelian reductions, construct
a source for the self-similar exterior field, and compare these
solutions with earlier Yang--Mills waves.

\emph{Discussion 1: Potentials and Abelian reductions.}
The potentials below retain the notation and gauge choices of the
preceding sections. Abelianity means the existence of a covariantly
constant color direction satisfying Eq.~\eqref{eq:III-U1criterion}.

For the five-parameter ansatz,
\begin{align}
 \varphi&=\alpha_1\Sigma_x,\qquad A_x=0,\qquad
 A_z=\alpha_2\Sigma_x,\nonumber\\
 A_y&=(\alpha_3+\alpha_5\cos\theta)\sigma_z
                      +\alpha_4\sin\theta\,\Sigma_y.
 \label{eq:VII-five-potentials}
\end{align}
The allowed coefficients are listed in Eq.~\eqref{eq:III-branches}
and the paragraph following it. All nonzero-field branches have
$\alpha_1=\alpha_2$, $\omega=ck$, and are $U(1)$ embeddings.
The zero-field branches are pure gauge.

The first transverse family has
\begin{align}
 A_x&=0,\qquad A_y=(c_0+c_4x)\sigma_z,\nonumber\\
 \varphi=A_z&=a_1\cos(\lambda y)\sigma_x
                         +a_2\sin(\lambda y)\sigma_y,
 \label{eq:VII-I-potentials}
\end{align}
where $a_1=(h_++h_-)/2$ and $a_2=(h_+-h_-)/2$.
For $c_4\ne0$, an Abelian reduction requires $\varphi=0$.
For $c_4=0$ and $\lambda\ne0$, the additional Abelian cases are
\begin{gather}
 c_0=0,\qquad a_1\equiv0\ \hbox{or}\ a_2\equiv0,\nonumber\\
 c_0=-\frac{\varepsilon\lambda}{2g},\qquad
 a_2=\varepsilon a_1,\qquad \partial_x^2a_1=0,
 \label{eq:VII-I-Abelian}
\end{gather}
with $\varepsilon=\pm1$. The choice $\varphi=0$ is Abelian for
arbitrary $c_0,c_4$. At $\lambda=0$, $a_2$ drops out of the
potential; the Abelian cases are $a_1=0$ or $c_0=c_4=0$.

For the second transverse family, $R=c_0+c_4x$ and
\begin{align}
 A_x&=0,\nonumber\\
 A_y&=R\sin\chi[-\sin(\lambda y)\sigma_x
                       +\varepsilon\cos(\lambda y)\sigma_y]
                       \nonumber\\
    &\quad+\left(R\cos\chi-\frac{\varepsilon\lambda}{2g}\right)
                       \sigma_z,\nonumber\\
 \varphi=A_z&=\left(\frac{\varepsilon k}{2g}\partial_\theta\chi+h\right)
              [\cos(\lambda y)\sigma_x
                       +\varepsilon\sin(\lambda y)\sigma_y].
 \label{eq:VII-II-potentials}
\end{align}
This family is Abelian for $h=0$ or $R\equiv0$.
For nonzero $R$ and $h$, the field commutator in
Eq.~\eqref{eq:IV-II-commutator} excludes a $U(1)$ reduction.

The remaining single-color solutions, for $\lambda\ne0$, have
$A_x=0$, $A_z=\varphi$, and either
\begin{align}
 \varphi&=\left(\frac{k}{\lambda}\partial_\theta c_1+h\right)
                      \cos(\lambda y)\sigma_x,\nonumber\\
 A_y&=-c_1\sin(\lambda y)\sigma_x,
 \label{eq:VII-single-x}
\end{align}
or
\begin{align}
 \varphi&=\left(\frac{k}{\lambda}\partial_\theta c_2+h\right)
                      \sin(\lambda y)\sigma_y,\nonumber\\
 A_y&=c_2\cos(\lambda y)\sigma_y.
 \label{eq:VII-single-y}
\end{align}
Here $h$ is given by Eq.~\eqref{eq:IV-single-color-mode}.
Each connection lies along one fixed Pauli matrix and is Abelian.
The flat connections, including the Painlev\'e III parametrization
in Eq.~\eqref{eq:IV-flat-coefficients}, have zero field strength
and are locally gauge-equivalent to the vacuum.

The self-similar family is
\begin{align}
 A_x&=-\frac{\sqrt2}{2gx}\sigma_z,\qquad
 A_y=-\frac{\sqrt6}{2gx}\sigma_x,\nonumber\\
 \varphi=A_z&=\Phi-\frac{3\sqrt2(ct+z)}{4gx^2}\sigma_z,
 \label{eq:VII-self-potentials}
\end{align}
with $\Phi$ given by Eq.~\eqref{eq:V-complete-profile}.
Here $[E_z,B_z]\ne0$ for every choice of the six profiles, so the
whole family is irreducible.

\emph{Discussion 2: Singularity and source.}
The self-similar field has a singular plane at $x=0$ for every
choice of the six profiles, as shown by
Eqs.~\eqref{eq:V-nonabelian} and \eqref{eq:V-energy}.
A source-supported exterior can be constructed by retaining
$x\ge a>0$ and joining it to a reflected copy. With $r=a+|x|$,
the joined potentials are
\begin{align}
 \widetilde\varphi&=\varphi(r),\qquad
 \widetilde A_z=A_z(r),\nonumber\\
 \widetilde A_y&=A_y(r),\qquad
 \widetilde A_x={\rm sgn}(x)A_x(r).
 \label{eq:VII-gluing}
\end{align}
All other arguments are unchanged. Each side satisfies the
vacuum equations. The tangential potentials are continuous,
so the field strength has a finite jump but no delta function.
Integrating $D_\mu F^{\mu\nu}=J^\nu$ across $x=0$ gives
\begin{gather}
 J^\nu=K^\nu\delta(x),\qquad K^\nu=[F^{x\nu}]_{0^-}^{0^+},
 \nonumber\\
 (K^0,K^x,K^y,K^z)
 =2\bigl(E_x(a),0,-B_z(a),E_x(a)\bigr).
 \label{eq:VII-sheet-current}
\end{gather}
The vacuum equations on the two sides imply $D_\nu K^\nu=0$.
For $\Phi=0$, the sheet supplies power per unit area
$2\mathcal S_x(a)=36c(ct+z)/(g^2a^5)$: it emits energy for
$ct+z>0$ and absorbs it for $ct+z<0$.

The background energy per unit area is
\begin{equation}
 \int_{-\infty}^{\infty}\widetilde{\mathcal E}\,{\rm d}x
 =\frac6{g^2a^3}+\frac{9(ct+z)^2}{g^2a^5}.
 \label{eq:VII-sheet-energy}
\end{equation}
It is finite for $a>0$, although the infinite sheet has infinite
total energy. The modes $f_1,f_5,f_6$ preserve transverse
convergence; an exterior extending to infinity requires
$f_2=f_3=f_4=0$. Both the current and stored energy diverge as
$a\to0$. Thus the construction supplies an exact exterior field
with a finite cutoff, rather than a finite-strength source for
the original singular plane.

\emph{Discussion 3: Comparison with earlier solutions.}
The principal differences are summarized in Table~\ref{tab:VI-comparison}.

\begingroup
\setlength{\intextsep}{8pt}
\begin{table}[H]
\caption{Propagation and spatial structure of Yang--Mills waves.}
\label{tab:VI-comparison}
\small
\renewcommand{\arraystretch}{1.12}
\setlength{\tabcolsep}{3pt}
\begin{tabular}{@{}p{0.30\columnwidth}p{0.66\columnwidth}@{}}
\hline\hline
{\raggedright Solutions\par} & {\raggedright Main features\par} \\
\hline
{\raggedright Ishida--Hosoya; Baseyan et al. \cite{IshidaHosoya1975,Baseyan1979}\par} & {\raggedright Nonlinear periodic waves with timelike wave vectors; Jacobi elliptic profiles.\par} \\
{\raggedright Kovacs--Lo: lightlike waves \cite{KovacsLo1979}\par} & {\raggedright Complex self-dual or anti-self-dual fields; arbitrary light-speed profiles; vanishing energy--momentum tensor.\par} \\
{\raggedright Coleman; Oh--Teh I \cite{Coleman1977,OhTeh1985,Cariglia2004}\par} & {\raggedright Light-speed profiles; pure-radiation flux; transverse harmonic profiles.\par} \\
{\raggedright Oh--Teh II \cite{OhTeh1985}\par} & {\raggedright Nonlinear periodic waves with phase velocity not fixed to $c$; regular and singular branches.\par} \\
{\raggedright Rabinowitch \cite{Rabinowitch2025,Rabinowitch2026}\par} & {\raggedright Progressive waves along $z$; exterior fields of sources near the axis.\par} \\
{\raggedright Present transverse families\par} & {\raggedright Arbitrary light-speed profiles; longitudinal magnetic field; $\mathcal S_z=c(\mathcal E-c_4^2)$.\par} \\
{\raggedright Present self-similar family\par} & {\raggedright Light-speed profiles on a $ct+z$-dependent background; longitudinal fields and a singular plane.\par} \\
\hline\hline
\end{tabular}
\end{table}
\endgroup
\par

For the transverse families, the longitudinal magnetic energy
$c_4^2$ does not contribute to $\mathcal S_z/c$. Light-speed
phase propagation therefore need not imply pure-radiation energy
transport. The self-similar family also has transverse energy
flux, supplied by the planar source constructed above. Its
source geometry differs from the axial exterior fields of
Refs.~\cite{Rabinowitch2025,Rabinowitch2026}.

The relevant nonexistence results have more restrictive
hypotheses. Basler and H\"adicke exclude genuinely non-Abelian
SU(2) fields with the full plane-wave symmetry group $G_5$
\cite{BaslerHadicke1984,AdamoIlderton2019}; planar phase fronts
alone do not impose that symmetry. Coleman's theorem concerns
nonsingular, finite-energy, source-free configurations that do
not radiate to spatial infinity \cite{ColemanGlueballs1977}.
The present fields are spatially extended or singular, and a
compact phase profile alone does not make them isolated
finite-energy configurations.

\section{Conclusions and Outlook}\label{sec:VII}

Our construction provides new analytic models for non-Abelian waves in SU(2) theory. The two transverse wave families and the finite-amplitude waves on a self-similar background show that arbitrary wave profiles can propagate at the speed of light while retaining a genuinely non-Abelian field structure. For the self-similar family with a singular plane, the surface-source construction further determines the charge and current needed to sustain propagation in the exterior region, allowing these solutions to be studied as exterior fields of sources. This also offers a possible physical interpretation of the solutions in terms of their sources and backgrounds.

Our ansatz construction starts from a pure-gauge field and proceeds by examining its structure and progressively enlarging the ansatz. This approach draws on the general observation that pure-gauge structures may occur as special cases of more general nontrivial configurations. Similar methods may also be useful for constructing other ans\"atze. In particular, there is room to develop the present ansatz further. In the extended equations, the transverse background can be determined before solving for the propagating part, whose equation remains linear on a given background. The self-similar background is one specific application of this method. Future work can use the same reduced equations to seek more general transverse backgrounds, examining how regularity and asymptotic decay constrain the wave profiles and whether genuinely non-Abelian waves can be localized transversely.

These solutions also provide concrete backgrounds for studying the stability of finite-amplitude non-Abelian waves. For the transverse families, one may begin with the magnetic background in the absence of a propagating wave, and then compare the evolution of perturbations after adding periodic waves or pulses, examining the respective effects of the propagating wave and the longitudinal magnetic field on stability. One could consider perturbations that modify the original transverse distributions and color structure, and combine this analysis with numerical evolution to study the nonlinear development of growing modes. Such analyses could identify which propagating states persist under small perturbations, providing a dynamical basis for their use in subsequent background-field studies.

For the self-similar family, a more direct physical question is how the constructed sources might be realized by matter. The surface charge and current obtained here specify the matching conditions between the source and the exterior field. One could introduce a layer of color-charged matter of finite thickness and study its self-consistent evolution with the exterior field, as well as whether the surface-source description is recovered in the thin-layer limit. This may provide a broader physical setting for the self-similar family and yield exact examples of sources and wave propagation in non-Abelian fields.
\begin{acknowledgments}
We thank Sayanjit Banerjee for pointing out that the solution labeled Family~II in the first arXiv version of this work is gauge-equivalent to an embedded $U(1)$ configuration.

This work is supported by the Quantum Science and Technology-National Science and Technology Major Project (Grant No. 2024ZD0301000), and the National Natural Science Foundation of China (Grant No. 12275136).
\end{acknowledgments}

\end{document}